\documentclass[useAMS,usenatbib,usegraphicx]{mn2e}

\usepackage{fixltx2e}[1999/12/01] 

\usepackage{aas_macros}     
\usepackage[dvips]{color}      
\newcommand{\lcdm}{$\Lambda$CDM } 
\renewcommand{\v}[1]{\bmath{#1}}      
\newcommand{\mat}[1]{\mathbfss{#1}}   
\newcommand{\dotprod}{\bmath{\cdot}}
\newcommand{\crossprod}{\bmath{\times}}
\newcommand{\rhocrit}{ \rho_\rmn{crit} }
\newcommand{\Omegatot}{\Omega_\rmn{tot}}
\newcommand{\Omegab}{\Omega_\rmn{b}}
\newcommand{\OmegaM}{\Omega_\rmn{M}}

\newcommand{\OmegaL}{\Omega_\Lambda}
\newcommand{\Msol}{\rmn{M_{\sun}}}   
\newcommand{\Mh}{M_\rmn{h}}  
\newcommand{\Npart}{N_{\rmn{part}} }
\newcommand{\Np}{ N_\rmn{p} }

\newcommand{\Nsel}{ N_\rmn{sel} }
\renewcommand{\mp}{ m_\rmn{p} }

\newcommand{\Rvir}{R_\rmn{vir}}
\newcommand{\Lbox}{ L_{\rmn{box}} }
\newcommand{\siglg}{\sigma_\rmn{lg}}
\newcommand{\lambmed}{\lambda_\rmn{med}}

\newcommand{\Mpc}{\rmn{Mpc}}            
\newcommand{\kpc}{\rmn{kpc}}            
\newcommand{\kms}{\rmn{km\,s^{-1}}}     
\newcommand{\Hunit}{ \, \kms \, \Mpc^{-1}} 
\newcommand{\munit}{ \, h^{-1} \Msol }     
\newcommand{\lunit}{ \, h^{-1} \Mpc }      
\newcommand{\klunit}{\, h^{-1} \kpc }      

\title[Spin of dark matter haloes]{The spin and shape of
  dark matter haloes in the Millennium simulation of a $\Lambda$CDM universe}

\author[Bett et al.]{Philip Bett$^{1,3}$, Vincent Eke$^{1}$, 
  Carlos S. Frenk$^{1}$, Adrian Jenkins$^{1}$, John Helly$^{1}$,
  \newauthor and Julio Navarro$^{2}$\\ 
  $^{1}$Institute for Computational Cosmology, University of Durham,
  South Road, Durham, DH1 3LE, UK\\
  $^{2}$Department of Physics and Astronomy, University of Victoria,
  Victoria, BC, V8P 5C2, Canada\\
  $^{3}$P.E.Bett@durham.ac.uk }

\begin{document}

\date{\today}

\pagerange{\pageref{firstpage}--\pageref{lastpage}} \pubyear{2006}

\maketitle

\label{firstpage}

\begin{abstract} 
We investigate the spins and shapes of over a million dark matter
haloes identified at $z=0$ in the Millennium simulation. Our sample
spans halo masses ranging from dwarf galaxies to rich galaxy clusters.
The very large dynamic range of this \lcdm cosmological simulation
enables the distribution of spins and shapes and their variation with
halo mass and environment to be characterised with unprecedented
precision. We compare results for haloes identified using three
different algorithms, and investigate (and remove) biases in the
estimate of angular momentum introduced by both the algorithm itself
and by numerical effects. We introduce a novel halo definition called
the TREE halo, based on the branches of the halo merger trees, which
is more appropriate for comparison with real astronomical objects than
the traditional ``friends-of-friends'' and ``spherical overdensity''
algorithms.  We find that for this many objects, the traditional
lognormal function is no longer an adequate description of the
distribution, $P(\lambda)$, of the dimensionless spin parameter
$\lambda$, and we provide a different function that gives a better fit
for TREE and spherical overdensity haloes.  The variation of spin with
halo mass is weak but detectable, although the trend depends strongly
on the halo definition used.  For the entire population of haloes, we
find median values of $\lambmed=0.0367$--$0.0429$, depending on the
definition of a halo.  The haloes exhibit a range of shapes, with a
preference for prolateness over oblateness.  More massive haloes tend
to be less spherical and more prolate.  We find that the more
spherical haloes have less coherent rotation in the median, and those
closest to spherical have a spin independent of mass ($\lambmed
\approx 0.033$).  The most massive have a spin independent of shape
($\lambmed \approx 0.032$).  The majority of haloes have their angular
momentum vector aligned with their minor axis and perpendicular to
their major axis.  We find a general trend for higher spin haloes to
be more clustered, with a stronger effect for more massive haloes.
For galaxy cluster haloes, this can be larger than a factor of $\sim
2$.
\end{abstract}


\section{Introduction}
The formation of galaxies is intimately linked
to the acquisition and distribution of angular momentum. In the
current cosmological paradigm, the inflationary \lcdm model, cosmic
structures grow hierarchically. Dark matter haloes form by the
dissipationless gravitational collapse of material associated with
peaks in the primordial density fluctuation field, growing, through
mergers and smooth accretion, into objects with a wide range of masses
at the present day. Galaxies form when baryons cool and condense near
the centre of these haloes (\citealt{1978MNRAS.183..341W},
\citealt{1991ApJ...379...52W}).  They undergo mergers and tidal
interactions along with their haloes, giving rise to the rich spectrum
of galaxy types and environments that we see today.

Understanding the generation and evolution of the angular momentum of
dark matter haloes is a prerequisite for understanding the
angular momentum and morphology of galaxies. For example, the
distribution of halo spins is a basic input to models of galaxy
formation (e.g. \citealt{1998MNRAS.295..319M},
\citealt{1998ApJ...507..601V}, \citealt{2000MNRAS.319..168C}).  The
early evolution of the angular momentum of a density perturbation is
adequately described by the linear tidal torque theory (see e.g.
\citealt{1949pca..conf..195H}, \citealt{1969ApJ...155..393P},
\citealt{1970Ap......6..320D}, \citealt{1984ApJ...286...38W},
\citealt{1996MNRAS.282..436C}, \citealt{2000ApJ...532L...5L}).
However, as work such as that of \cite{1984ApJ...286...38W} and
\cite{2002MNRAS.332..325P} has shown, the non-linear effects inherent
in the formation of large-scale structure lead to large quantitative
disagreements between the predictions of the tidal torque theory and
the angular momenta found in $N$-body simulations of dark matter haloes.

$N$-body simulations provide the way to progress beyond the linear regime.  As
computing power has improved, so has the scale and resolution of
simulations.  Very early numerical studies of the angular
momentum of ``proto-galaxies'' were performed by
\cite{1971A&A....11..377P} (with $\Np\sim 100$ particles) and
\cite{1979MNRAS.186..133E} ($\Np=1000$), and led the way to the
analysis of the spins and shapes of CDM haloes in more sophisticated
simulations (\citealt{1985ApJ...292..371D, 1987ApJ...319..575B,
1988ApJ...327..507F}; all with $\Np=32\,768$).
\cite{1992ApJ...399..405W} used a much larger simulation ($\Np\sim
10^6$) and focused particularly on the details of the distributions of
halo spins and shapes, and their relationship through the alignment of
the halo angular momentum vector.  \cite{1996MNRAS.281..716C} also
investigated the shapes and spins of dark matter haloes, in addition to
various other aspects of halo structure. 

An early study by \cite{1987ApJ...319..575B} examined the relationship
between spin and the spatial clustering of haloes, as measured by the
two-point correlation function. Later, \cite{1999MNRAS.302..111L}
examined the environmental dependence of halo properties, and found no
correlations with halo spin. More recently, \cite{2002A&A...395....1F}
have carried out a mark correlation function analysis to investigate
how spin varies with halo pair-separation.  They found that
neighbouring cluster pairs tend to have higher spins than the average.

Recent years have seen a large amount of work on the analysis of
haloes in \lcdm simulations.  Halo shapes and their variation with
mass were investigated by \cite{2002sgdh.conf..109B},
\cite{2005ApJ...629..781K}, \cite{2006ApJ...646..815S} and
\cite{2006MNRAS.367.1781A}. In agreement with previous studies, halo
spin was found to vary little, if at all, with halo mass. The
relationship between halo shape and spin was investigated by
\cite{2005ApJ...627..647B}, \cite{2005ApJ...634...51A},
\cite{2006ApJ...646..815S} and \cite{2006EAS....20...25G}.  While
this paper was being completed, independent analysis of halo
properties, investigating halo concentrations, spins and shapes in a
series of simulations, was posted by \cite{2006astro.ph..8157M}.

Using the the 10-billion particle Millennium simulation of the
evolution of dark matter in the \lcdm cosmology
(\citealt{2005Natur.435..629S}), we re-examine, in this paper, some of
the shape and spin properties of dark matter haloes previously
considered. Our analysis improves upon earlier work because the
millions of haloes that formed in this simulation provide
unprecedented statistical power. This allows us, for example, to
quantify the distribution of halo spins and the relationship between
spin, halo mass and shape with a precision that has not hitherto been
possible.  Unlike previous work, we consider different ways to
identify haloes in the simulations; it turns out that the details of
halo definition and selection can have a strong impact on the results.
Finally we investigate how halo clustering depends on spin and shape
for haloes of different masses.

This paper is structured as follows.  Section \ref{s:dmh} provides a
description of the Millennium simulation itself, and the various halo
properties we shall be investigating. Section \ref{s:halocats}
describes the construction of the catalogues whose haloes we
investigate, including the group-finding algorithms and halo selection
criteria.  These we use to remove haloes whose properties are
unreliable or biased, due to both numerical effects and the
group-finding algorithms themselves.  The main results of this paper
are presented in \S\ref{s:results}, where we describe the distribution
of halo spins as a function of mass and shape, and examine its effect
on halo clustering.  Our conclusions are presented in \S\ref{s:conc}.
Finally, the Appendix shows various examples of haloes that illustrate
the effects of the group-finders.

\section{Dark Matter Halo Properties in the Millennium simulation}\label{s:dmh}
\subsection{The simulation}
The Millennium simulation is described by
\cite{2005Natur.435..629S}. It followed the evolution of 10 billion
dark matter particles in the \lcdm model, the standard paradigm of
modern cosmology.  This is is strongly favoured by measurements of the
temperature anisotropies in the microwave background radiation
\citep{2003ApJS..148..175S, 2006astro.ph..3449S} and by measurements
of the clustering of galaxies \citep{2002MNRAS.337.1068P,
2004PhRvD..69j3501T, 2006MNRAS.366..189S}.  For reference, the key
parameters of the simulation are listed in Table~\ref{simparams}.

\begin{table}
  \begin{center}
    \begin{tabular}{c|c|c|c|c|c}\hline
      $\OmegaL$ & $\OmegaM$ & $\Omegab$ & $h$ & $n$ & $\sigma_8$ \\
        $0.75$     & $0.25$ & $0.045$ & $0.73$ & $1.00$ & $0.9$ \\
      \hline\hline
      $\Lbox$        & \multicolumn{2}{c|}{$\Npart$} & 
          \multicolumn{2}{c|}{$\mp$}    & $\eta$ \\
      $500\lunit$      & \multicolumn{2}{c|}{$2160^3\approx 10^{10}$} &
           \multicolumn{2}{c|}{$8.6\times 10^8\munit$}    & $5\klunit$ \\
      \hline
    \end{tabular}
    \caption{Cosmological and simulation parameters for the Millennium
      Run. The first row describes the cosmology used.  It
      gives the density parameters $\Omega_i:=
      \frac{\rho_i}{\rhocrit}$, where the critical density
      $\rhocrit:=\frac{3H_0^2}{8\pi G}$ (with
      $\rho_\Lambda:=\frac{\Lambda}{8\pi G}$ and $\Omegatot = \OmegaL
      +\OmegaM =1$), the Hubble parameter where $H_0=100h(\!\Hunit)$,
      the spectral index, $n$, and $\sigma_8$, the linear-theory power spectrum
      variance in spheres of radius $8\lunit$.  The
      second row gives the simulation parameters: the length of the side
      of the simulation cube, the number of particles, the
      resulting particle mass and the gravitational softening length.
      For further details, see Springel et al. (2005).}
    \label{simparams}
  \end{center}
\end{table}

The simulation gives the positions and velocities of all particles at
each ``snapshot'' in time; $64$ snapshots are stored from the initial
redshift ($z=127$) to the present, enabling redshift-dependent
statistics to be studied.

\subsection{Halo properties}\label{s:properties}
For each halo-finding algorithm we consider in this paper, we compute a
range of halo properties.  We shall discuss these quantities here, and
defer the details of the halo-finding and halo-selection algorithms to
the next section.

Much of this work concentrates on the dimensionless spin parameter
$\lambda$, introduced by
\cite{1969ApJ...155..393P,1971A&A....11..377P}. This is defined as:
\begin{equation}
  \lambda := \frac{J|E|^{1/2}}{G\Mh^{5/2}} \equiv \frac{j|E|^{1/2}}{G\Mh^{3/2}}
  \label{e:spin}
\end{equation}
where $\Mh$ is the halo mass, $J$ is the magnitude of the angular
momentum vector $\v{J}$ (and $j$ is the specific angular momentum),
$E$ is the total energy, and $G$ is Newton's gravitational constant.
It is important to note that $\lambda$ is defined for any object which
has a well-defined $j$, $E$ and $\Mh$; all these quantities are
conserved for an isolated system, virialised or not. In reality, none
of the haloes in cosmological simulations are completely isolated,
leading to ambiguities in the definition. This means that for the
$\lambda$ of a halo to be useful, it is the definition of \emph{halo}
that requires the most care (and conditions such as virialisation; see
section \ref{s:cleaning}), not the definition of $\lambda$.  

The meaning of $\lambda$ is therefore best understood by considering
an isolated, virialised, spherical system.  The spin parameter can be
seen to be a measure of the amount of coherent rotation in a system
compared to random motions.  For a spherical object, it is
approximately the ratio of its own angular velocity to the angular
velocity needed for it to be supported against gravity solely by
rotation \citep[see e.g.][]{1993sfu..book.....P}.

The specific angular momentum $\v{j}$ and kinetic energy $T$ of each
halo containing $N_p$ particles are given by:
\begin{eqnarray}
  \v{j}   & = & \frac{1}{\Np} \sum_{i=1}^{\Np} \v{r}_i\crossprod \v{v}_i \\
        T & = & \frac{1}{2} \Mh \sum_{i=1}^{\Np}\v{v}_i^2
\end{eqnarray}
where $\v{r}_i$ is the position vector of particle $i$ relative to the
halo centre, and $\v{v}_i$ is its velocity relative to the halo centre
of momentum.

The halo potential energy, $U$, is calculated using all halo particles
if $\Mh\leq 1000\mp$, and is rescaled up from that of $1000$
randomly-sampled particles otherwise.  The potential is that used in
the simulation itself:
\begin{equation}
  U = \left(\frac{\Np^2-\Np}{\Nsel^2-\Nsel}\right)
         \left(\frac{-G \mp^2}{\eta}\right)
         \sum_{i=1}^{\Nsel-1} \sum_{j=i+1}^{\Nsel} -W_2(r_{ij}/\eta)
\end{equation}
where $\Nsel$ is the number of selected particles ($\Nsel\leq
1000$), $\eta$ is the softening length (see Table \ref{simparams}),
$r_{ij}$ is the magnitude of the separation vector between the $i$th
and $j$th particles in the halo, and the softening kernel
\citep[see][]{2001NewA....6...79S} is:
\begin{equation}
  W_2(u) =
     \left\{
        \begin{array}{ll}
           \frac{16}{3}u^2 -\frac{48}{5}u^4 +\frac{32}{5}u^5 -\frac{14}{5}, &
	   0\leq u \leq\frac{1}{2}, \\

           \frac{1}{15u} +\frac{32}{3}u^2 -16u^3 +\frac{48}{5}u^4 \\
	   \;\;\; -\frac{32}{15}u^5 -\frac{16}{5}, &
	   \frac{1}{2}\leq u\leq1, \\

           -\frac{1}{u}, &  u\geq 1 \\
	   
        \end{array}
     \right.
\end{equation}

A halo's shape is derived from its mass distribution, which we
characterise using the inertia tensor, $\mat{I}$.  This relates
angular momentum $\v{J}$ and angular velocity $\v{\omega}$ through
$\v{J}=\mat{I}\v{\omega}$, and is formed from the following
components:
\begin{equation}
  I_{\alpha\beta} := \sum^{\Np}_{i=1} m_i \left( \v{r}_i^2
  \delta_{\alpha\beta} - r_{i,\alpha}r_{i,\beta} \right) 
  \label{e:inertia}
\end{equation}
where $\v{r}_i$ is the position vector of the $i$th particle, $\alpha$
and $\beta$ are are the tensor indices with values of 1, 2 or 3, and
$\delta_{\alpha\beta}$ is the Kronecker delta.  The process of
diagonalising $\mat{I}$ is equivalent to rotating the coordinate
system to find a set of axes in which a torque about one does not
induce a rotation about another; i.e. such that $\v{J}$ is parallel to
$\v{\omega}$.  These axes then describe a hypothetical uniform
ellipsoid whose axes $a\ge b\ge c$ are those of the halo itself:
\begin{equation}
  \mat{I} = \frac{1}{5} \Mh \left( \begin{array}{ccc}
    b^2+c^2 & 0       & 0      \\
    0       & a^2+c^2 & 0      \\
    0       & 0       & a^2+b^2
  \end{array} \right)
  \label{e:inertiadiag}
\end{equation}
The eigenvalues are the moments of inertia $\mathcal{I}$ for rotation
about that axis.  For example, rotation about the semimajor axis
$\v{a}$ has the moment of inertia
$\mathcal{I}_a=\frac{1}{5}\Mh(b^2+c^2)$; note that $\mathcal{I}_a \le
\mathcal{I}_b \le \mathcal{I}_c$.  These eigenvalues can then be
combined to find the relative axis lengths, e.g.
\begin{equation}
  a = \sqrt{\frac{5}{2\Mh}\left(-\mathcal{I}_a + \mathcal{I}_b +
  \mathcal{I}_c  \right) }
  \label{e:axes}
\end{equation}

The axis vectors are given directly by the corresponding eigenvectors,
so that, for example, rotation about the $\v{a}$-axis of the ellipsoid
(with the longest length $a$) has the smallest moment of inertia,
$\mathcal{I}_a$.

Much of the literature on halo shapes uses the following description
of the mass distribution, confusingly also calling it the inertia
tensor (see, e.g. \citealt{1996MNRAS.281..716C},
\citealt{2005ApJ...618....1H}, \citealt{2005ApJ...627..647B},
\citealt{2006ApJ...646..815S}):
\begin{equation}
  \mathcal{M}_{\alpha\beta}
  = \frac{1}{\Np}\sum^{\Np}_{i=1} r_{i,\alpha}r_{i,\beta}
\end{equation}
The results are entirely equivalent: if one diagonalises this matrix,
then $a$, $b$ and $c$ can be found as just the square roots of the
eigenvalues, and the eigenvectors again give the axis vectors.

Once the halo's principal axes have been found, relationships between
the axes and between the shape and other properties such as spin can
be examined in terms of the axis ratios $p:=c/b$, $q:=b/a$ and
$s:=c/a$.  The minor-to-major axis ratio $s$ is a useful measure of
the sphericity of the system, but does not specify in what way a halo
might be aspherical.  For this, we can use the triaxiality parameter
introduced by \cite{1991ApJ...383..112F}:
\begin{equation}
  \mathcal{T} = \frac{a^2-b^2}{a^2-c^2}
\end{equation}
This measures whether a halo is prolate ($\mathcal{T}=1$) or oblate
($\mathcal{T}=0$), but it does not quantify how aspherical a halo is.

\section{The halo catalogues}\label{s:halocats}
\subsection{Groupfinder algorithms}
The problem of how best to identify groups of particles within
$N$-body simulations is a ubiquitous feature in studies of dark-matter
halo properties.  Many solutions have been found to this problem,
typically involving first finding candidate-halo centres, followed by
an iterative scheme to shrink or grow the halo according to criteria
involving the binding energy or overdensity.  Such algorithms include
Bound Density Maxima (BDM, \citealt{1999ApJ...516..530K}), Spline
Kernel Interpolative DENMAX (SKID, \citealt{1997ApJ...477....8W}), the
AMIGA Halo Finder (AHF, \citealt{2004MNRAS.351..399G}), and
\textsc{Subfind} \citep{2001MNRAS.328..726S}, which we use below.

The very simple `friends-of-friends' group-finder (FOF,
\citealt{1985ApJ...292..371D}) was run on-the-fly, during the
Millennium simulation run, with a linking length of
$s_0=b(\Lbox^3/\Npart)^{1/3}$ where $b=0.2$, to attempt to select
virialised structures in the particle distribution.  As is often the
case (e.g. \citealt{2002MNRAS.332..325P},
\citealt{2006ApJ...646..815S}, \citealt{2006astro.ph..8157M}), this
simple FOF catalogue forms the basis for the more sophisticated halo
definitions we use.

An enhanced version of the \textsc{Subfind} program
\citep{2005Natur.435..629S} was run on the data to identify self-bound
substructures within each FOF halo, which we then use to construct the
different group catalogues we investigate.  The \textsc{Subfind}
algorithm is essentially a two-step process.  The first task is to
identify subhalo candidates within each FOF halo.  This is done using
an adaptively-smoothed dark matter density field, effectively lowering
a density threshold and identifying the peaks that grow out of it.
The second stage consists of performing an iterative gravitational
unbinding procedure on the candidates, successively removing particles
that are not bound to the subhalo candidate.  For this purpose, the
potential energies are computed using a tree algorithm similar to that
used for the simulation itself.  The candidates that are left with at
least $20$ particles after this procedure are then subhaloes of the
parent halo.  The algorithm can and does identify subhaloes within
subhaloes.  It results in the FOF haloes typically consisting of a
hierarchy of self-bound structures (which are not necessarily bound to
each other), and a set of particles (referred to as ``fuzz'') that are
spatially linked to the halo but not part of any self-bound
(sub)structure.  The most massive ``subhalo'' (MMSH) typically
contains most of the mass of the corresponding FOF object, and so is
best regarded as the self-bound background halo itself, with the
remaining subhaloes as its substructure.

In addition to finding the bound structures within haloes,
\textsc{Subfind} also computes certain subhalo properties,
which are then stored in the subhalo catalogue files.  These include
the location of the potential minimum, the ID number of the most bound
particle, the mass (number of particles), and the half-mass radius.
\textsc{Subfind} also computes and stores additional data related to
each parent halo.  Starting at the potential minimum of the MMSH of a
halo, three radii are found: the first two are those where the density
within them drops below $200\rhocrit$ and $200\bar{\rho}$, to aid
comparison with other work. The third is the virial radius proper,
which uses the fitting formula of \cite{1998ApJ...495...80B} for
spherical top-hat collapse in a flat ($\OmegaL+\OmegaM=1$) cosmology
(see also \citealt{1996MNRAS.282..263E}):
\begin{equation}
  \frac{\rho}{\rhocrit}= 18\pi^2+82(\OmegaM(z)-1)-39(\OmegaM(z)-1)^2
\end{equation}
This gives $\rho/\rhocrit \approx 94$ at $z=0$.  Although these
properties are associated with sets of grouped particles,
\textsc{Subfind} does not restrict itself to these particles when
computing them.  

We will now describe the group catalogues whose halo properties we
have investigated, and how they are built from the results of the FOF
and \textsc{Subfind} algorithms.  A key point which is used for each
halo definition is that we take the halo centre to be at the
potential minimum of the MMSH.  For reference, Table~\ref{t:halocats}
gives a list of each halo catalogue we will discuss, and the number of
haloes they contain.

\begin{table}
  \begin{center}
    \begin{tabular}{lrl}
      \hline
      Halo catalogue & \multicolumn{2}{l}{Number of haloes} \\
      \hline
      FOF (without \textsc{Subfind})  &  $17\,709\,121$  & (`raw' catalogue)\\
      \hline
      FOFall    &  $15\,494\,624$ & ($87.5\%$ of FOF raw)\\
      FOFclean  &  $ 1\,332\,239$ & ($8.60\%$ of FOFall)\\
      \hline
      SOall     &  $15\,458\,379$ & ($99.8\%$ of FOFall)\\
      SOclean   &  $ 1\,239\,494$ & ($8.02\%$ of SOall)\\
      \hline
      TREEall   &  $17\,041\,498$ & ($110.0\%$ of FOFall)\\
      TREEclean &  $ 1\,503\,922$ & ($8.83\%$ of TREEall)\\
      \hline
    \end{tabular}
    \caption{Numbers of haloes in the halo catalogues defined in this
      paper.  The ``all'' catalogues are the unfiltered results of the
      groupfinder algorithms described in sections \ref{s:fofdef},
      \ref{s:sodef} and \ref{s:treedef}.  The ``clean'' catalogues
      have been filtered as discussed in sections \ref{s:removal} and
      \ref{s:numprobs}, with the quasi-equilibrium parameter $Q=0.5$
      and the particle-number limit $\Np\ge 300$.}
    \label{t:halocats}
  \end{center}
\end{table}

\subsubsection{FOF haloes}\label{s:fofdef}
The basic FOF algorithm, run as already described, results in
$17\,709\,121$ groups containing at least $20$ particles, i.e. of mass
$\ga 1.7\times 10^{10}\munit$, at redshift $z=0$.  The most massive
group contained $4\,386\,162$ particles ($\approx 3.8\times
10^{15}\munit$). Using this catalogue we can therefore identify dark
matter haloes over a range of more than 5 orders of magnitude in mass,
ranging from subgalactic clumps to the most massive clusters.

In practice, since we use the centre (potential minimum) of the MMSHs
as the centres of the haloes themselves, we only use FOF haloes for
which \textsc{Subfind} has found bound substructures.  Haloes without
substructure (and hence without an MMSH) are excluded from our base
FOF catalogue.  This has the effect of reducing the catalogue size by
$12.5\%$, preferentially at lower masses.  We shall refer to this
catalogue as FOFall, and we will not discuss the larger raw FOF
catalogue further.

\subsubsection{SO haloes}\label{s:sodef}
The properties calculated by \textsc{Subfind} make it possible to construct
a second halo catalogue, in which each halo consists of only the
particles within $\Rvir$ of the centre of the MMSH of the
corresponding FOF object (note that these particles do not have to be
members of the FOF halo).  Due to the way \textsc{Subfind} constructs
the MMSH, this yields haloes whose definition is similar to those from
a ``spherical overdensity'' algorithm \citep{1994MNRAS.271..676L}, so
we will refer to them as the SO haloes.  We do not impose a
lower limit on the number of particles comprising these objects; as
a result, haloes can be identified with masses $<20\mp$ when their
virial radii encompass fewer particles than their original FOF halo.
This is simply a consequence of the algorithm employed; later
examination of halo spins reveals the need for a much higher
particle-number limit, as discussed in full in
Section~\ref{s:numprobs}.  This halo definition is similar, but not
identical, to that used in \cite{2006astro.ph..8157M}.

\subsection{Better halo catalogues}\label{s:cleaning}
\subsubsection{Groupfinder problems}\label{s:groupprobs}
The FOF and SO group-finding algorithms have various well-documented
drawbacks.  Groupfinders such as SO which use the overdensity
contained within a spherical region tend to impose a more spherical
geometry on the resulting systems.  Although this is not a problem for
many objects, the algorithm can sometimes result in very
unnatural-looking structures.  An example is shown in figure
\ref{f:halo_SObad}.  This compares a massive FOF halo with the
corresponding SO object.  The centre of mass of the FOF system is well
separated from the minimum of the potential, and the halo is
significantly elongated.  This results, when growing a sphere around
the potential minimum to form the SO halo, in the virial overdensity
being reached sooner in one direction than another.  The SO halo
contains particles in low-density regions outside the FOF halo, and
has a sharp cutoff in another direction when the FOF halo continues.
The more `normal' haloes, in the background, are much less affected.

The problems associated with the FOF groupfinder can be more extreme,
and can affect a greater proportion of the haloes.  One of the most
commonly-cited problems (e.g. \citealt{1994ApJ...436..467G},
\citealt{1997NewA....2...91G}) is that well-resolved objects
identified using the FOF algorithm are often at risk of becoming
linked with neighbouring objects via tenuous bridges of particles.
Low-mass particle bridges are usually extremely transient structures,
being just a chance grouping of particles at that instant in time.
The joining of two (or more) otherwise unrelated objects of similar
mass in this way results in a very large velocity dispersion.
Examining the halo in velocity-space will clearly show the
multiple-object nature of the system.  An example of a halo formed
from objects joined by a tenuous bridge can be seen in
Fig. \ref{f:halo_bridging}.

Sometimes more massive haloes can be formed by the chaining together
of somewhat smaller objects that are undergoing mergers or close
flybys with their neighbours.  Their multi-object composition can
again be seen in velocity space as well as in real space, and although
their connections are likely to be less transitory than in the case of
a thin bridge, these objects are nevertheless well out of their
equilibrium state, and so are unhelpful when trying to characterise
the spins of typical dark matter haloes.  See Fig.
\ref{f:halo_largermulti} for an example of a larger multi-object halo.

A similar effect is that of velocity contamination of small objects
due to their proximity to more massive ones (see Fig.
\ref{f:halo_smallneighbour} for an example).  Just as particles can
form a bridge between passing haloes at the moment of the snapshot, so
an individual particle orbit can take it within the linking length of
a neighbouring halo, without forming enough of a bridge for the haloes
themselves to be joined.  The smaller halo will be contaminated by
these interloper particles, which will have a quite different mean
velocity to the halo's own particles.  This causes the mean velocity to be
shifted away from that of the `original' halo, and the resulting halo
to have a much larger velocity dispersion than expected for an object
of that mass.  The massive neighbouring object will have a much higher
velocity dispersion anyway, so will be unaffected by such effects.

\begin{figure}
  \includegraphics[width=84mm]{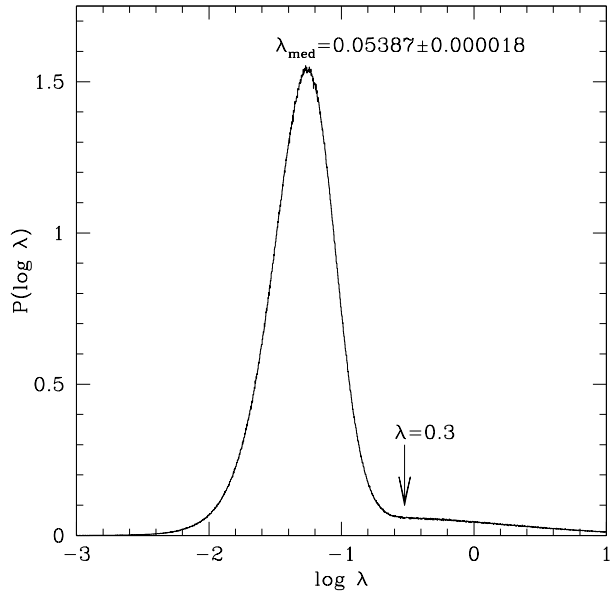}
  \caption{Histogram of the spin parameters from the basic FOFall halo
  catalogue of the Millennium Run, showing a long tail to high spins.
  The tail continues up to $\lambda\approx 680$, and there are over
  $900\,000$ objects with $\lambda\ge 0.3$ (marked on the graph).  The
  median spin of the distribution, $\lambmed$, is displayed with the
  uncertainty given by Eqn.~\ref{e:mederr}.  This demonstrates the
  need for more careful definition and selection of haloes.}
  \label{f:spinhist_FOF_all}
\end{figure}

\begin{figure}
  \includegraphics[width=84mm]{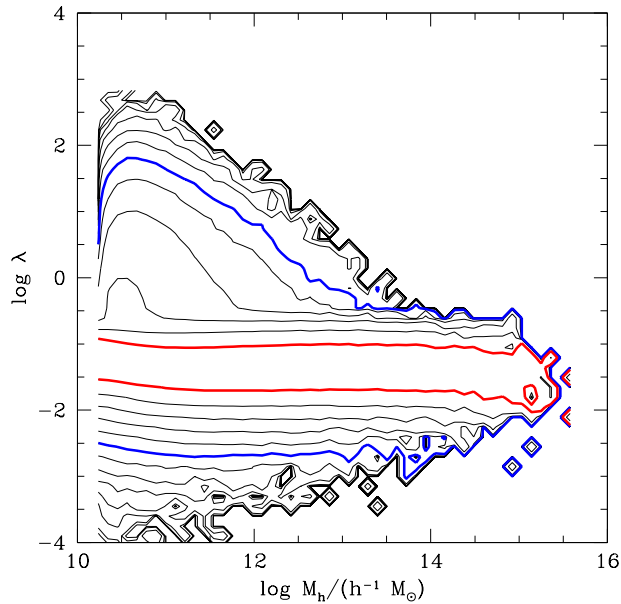}
  \caption{Halo spin as a function of halo mass for the FOFall halo
  catalogue.  The contours indicate the relative number density of
  haloes with that value of $\lambda(\Mh)$; that is, the haloes were
  binned onto a $50\times 50$ grid between the maximum and minimum
  values in $\log \Mh$ and $\log \lambda$, and the number of haloes in
  each grid cell was normalised by the number of haloes in that
  mass-bin, thus removing the effects of the halo mass function from
  the plot.  The contours are spaced logarithmically, with one contour
  for every factor of $10^{0.5}$ in halo number density.  The
  innermost bold contour (red) represents $10^{-1}$ (i.e. a tenth of
  the haloes in each mass bin), and the outer bold contour (blue)
  represents $10^{-3.5}$.  Notice the high-spin bulge, which extends
  over a large range of halo masses.  The results for SO haloes show a
  very similar distribution. }
  \label{f:fof_spinmass_all}
\end{figure}

To illustrate the effect that these problems have on the physical
properties we calculate, we show the spin distribution of the FOFall
catalogue in Fig. \ref{f:spinhist_FOF_all}.  It shows a long tail at
higher spins; there are $900\,748$ objects ($6\%$ of the catalogue)
with spin $\lambda\ge 0.3$. Fig. \ref{f:fof_spinmass_all} shows the
FOFall spin distribution as a function of halo mass, rescaled to
show the fractional number of objects at each mass so the trend of the
mass function is removed.  It shows that the high-spin tail comes from
objects over a large range of masses, and is therefore not due to
under-resolving groups.  These presumed-anomalous high spins come from
objects with high velocity dispersions for their masses, caused by
situations such as those described above.  This increases the haloes'
kinetic energies, $T$, leading to large values of $\lambda$.

These features are not usually seen in other published spin
distributions (e.g. \citealt{2001ApJ...557..616G},
\citealt{2002ApJ...581..799V}, \citealt{2005ApJ...634...51A},
\citealt{2006ApJ...638L..13T}, among many others), partly because more
advanced groupfinders and halo selection criteria are often used (as
we do below). However, the fact that we can see these artifacts so
clearly is because the Millennium Run gives us a vast number of
objects, over a wide range of masses.

For convenience, we shall refer to haloes suffering from the problems
described in this subsection as ``mis-defined'' haloes, as their
anomalously-high spins originate in how the haloes are defined by the
groupfinder algorithm in relation to their environment, rather than
any physical or numerical effect.

\subsubsection{A better groupfinder: The TREE haloes}\label{s:treedef}
As a third definition of halo, we use the `merger-tree' haloes
described by \cite{2006MNRAS.367.1039H}.  These are the $z=0$ objects
in a catalogue of halo merger trees constructed in the Millennium Run
(Helly et al., in preparation). These merger trees are similar to, but
distinct from those of \cite{2005Natur.435..629S} and
\cite{2005MNRAS.363L..66G}, who used different criteria for
identifying and tracking the haloes over time.  The merger-tree-halo
catalogue used here was designed with the needs of the $N$-body
GALFORM semi-analytic models in mind
(e.g. \citealt{2003MNRAS.338..903H}, \citealt{2006MNRAS.370..645B}).

The haloes in the TREE catalogue are formed from the \textsc{Subfind}
subhalo catalogue (based on the FOF haloes) in the following way.
Each halo is initially taken to consist solely of its constituent
subhaloes (i.e., is equivalent to the corresponding FOF halo, but
without the ``fuzz'').  Next, a ``splitting'' algorithm is applied, which
attempts to account for the linking of distinct, bound objects that
often occurs with FOF.  A subhalo can be split off from its parent
halo if it satisfies at least one of the following conditions: 1) The
distance between the subhalo's centre and the parent's centre is more
than twice the parent's half-mass radius; or 2) the subhalo has
retained at least $75\%$ of the mass it had at the last snapshot in
which it was an independent halo.  This second condition, which uses
the merger-tree data, is based on the idea that if the subhalo was
genuinely merging with the halo, it would be rapidly stripped of its
mass.  If however it was just a rapid encounter causing an artificial
link, then its mass would be retained.  If the subhalo is split off,
any other subhaloes that reside within twice \emph{its} half-mass
radius are also split off to become part of the new halo. We shall
refer to the catalogue of all haloes defined in this method as the
TREEall catalogue.

This definition of the TREE-halo catalogue alleviates the
groupfinder-based problems somewhat.  The unbound particles excluded
from the TREE haloes have, by definition, higher velocities than the
bound structures in the haloes.  Their removal therefore reveals the
more relaxed underlying haloes, with lower kinetic energies and hence
$\lambda$.  ``Interloper particle'' contamination, in particular, is
reduced by this feature.  The halo splitting algorithm also helps in
many cases by separating objects that have been spuriously linked by
bridges.

\begin{figure}
  \includegraphics[width=84mm]{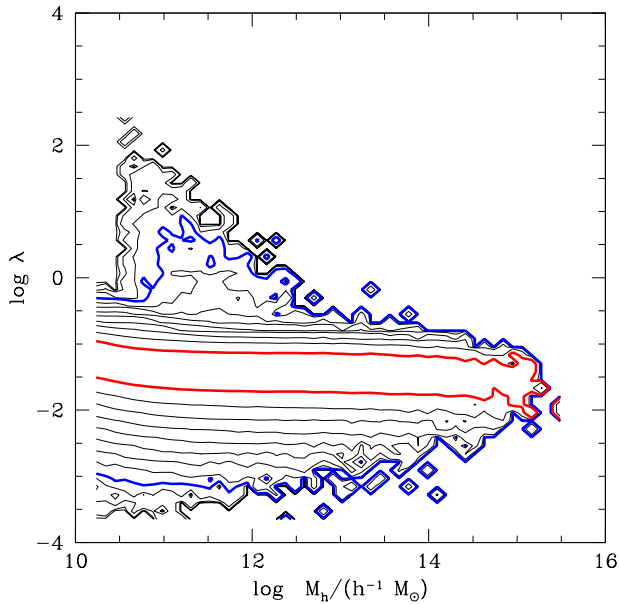}
  \caption{Spin as a function of halo mass for haloes in the TREEall
  catalogue. The contouring is as in Fig. \ref{f:fof_spinmass_all},
  i.e. in equal logarithmic steps of $10^{0.5}$, normalised to remove
  the mass function.  In this plot, the inner bold contour (red)
  represents $10^{-1}$ of the haloes at each mass, and the outer bold
  contour (blue) represents $10^{-5}$ of the haloes at each mass.  The
  merger-tree halo definition has moved many of the high-spin haloes
  visible in Fig. \ref{f:fof_spinmass_all} down into the main body of
  the distribution.}
  \label{f:tree_spinmass_all}
\end{figure}

The spin-mass distribution of the TREEall catalogue can be seen in
Fig. \ref{f:tree_spinmass_all}; although some of the high-spin objects
remain, the majority are now gone (there are sufficiently few high-spin
objects remaining that they are invisible in a plot of
$P(\log\lambda)$ such as Fig \ref{f:fof_spinmass_all}).  Investigating
the remaining high-spin objects reveals them to be victims of velocity
contamination from very massive neighbours.  They are somewhat special
cases however: for them not to have been rejected as fuzz, the
contaminants must be self-bound bodies, with $\geq 20$ particles each.
They must also be built on density maxima independent of that of the
host halo, in order for \textsc{Subfind} to have identified them as
separate subhaloes.  Furthermore, the interloping subhalo must be
within twice the half-mass radius of the halo, so that the tree
algorithm does not split it off.  A consequence of this is that the
resulting halo must consist of at least $40$ particles ($20$ for the
interloper subhalo and $20$ for the main halo), and this can be seen
in the offset low-mass cutoff for high-$\lambda$ haloes in Fig.
\ref{f:tree_spinmass_all}.

\subsubsection{A better halo catalogue: The QE criterion}\label{s:removal}
A relatively simple way of culling the remaining anomalous spin
objects is to remove those that are clearly out of equilibrium at the
moment of the snapshot.  This is not quite the same as selecting only
objects that are within a certain degree of true virialisation, since
we don't have the necessary time-resolution to determine if the system
properties are genuinely stationary: just as an object can appear to
be linked to another by a bridge that may exist only fleetingly, so a
halo could instantaneously have very similar energies to those of a
stationary system.  Therefore we will apply a cut in the instantaneous
`virial ratio' of halo energies, $2T/U+1$, and describe the objects
that meet this criterion as haloes in a \emph{quasi-equilibrium} (QE)
state.  This name avoids implying the zero time-derivative necessary
for the true virial ratio.

The question of where to make the cut in `virialisation'
(i.e. applying a QE limit) is a difficult one, because the decision
will always be somewhat arbitrary.  Since it is desirable to minimise
the effect of such arbitrariness, we concentrate on the effect of
applying a QE limit to the TREE haloes, since the merger-tree
algorithm has already removed many of the mis-defined haloes.

\begin{figure}
  \includegraphics[width=84mm]{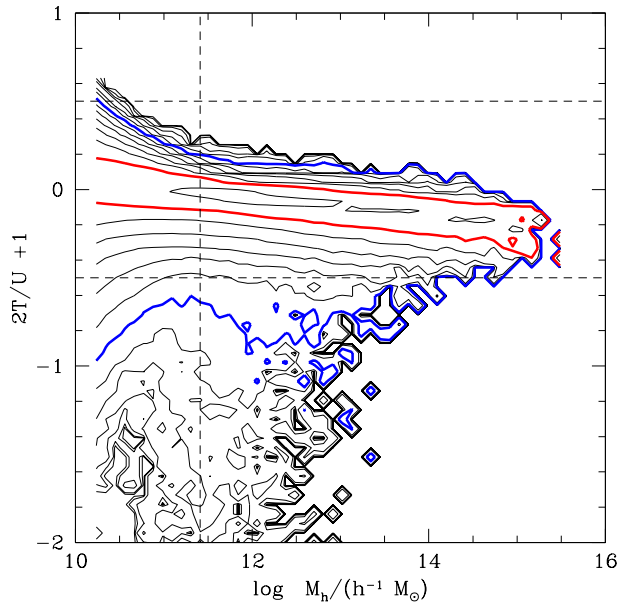}
  \caption{Contour plot of the instantaneous `virial ratio', $2T/U+1$,
    against halo mass for TREE haloes.  A virialised object has a
    value around zero, and a gravitationally bound object has value
    $>-1$.  The tail at low values (large $T$) extends down to
    $2T/U+1\approx -960$; there are $3733$ objects with $2T/U
    +1\le-1$.  The dashed lines show the QE limit of $Q=0.5$, and the
    lower particle-number limit of $\Np=300$.  The contouring is as in
    Fig.~\ref{f:fof_spinmass_all}, i.e. relative halo number density
    in equal logarithmic steps of size $10^{0.5}$.  The inner bold
    contour (red) represents $10^{-1}$ of the haloes at each mass, and
    the outer bold contour (blue) represents $10^{-4}$ of the haloes
    at each mass. The plots for FOF and SO groups are similar to
    this.}
  \label{f:virrat_mass_tree_all}
\end{figure}

\begin{figure}
  \includegraphics[width=84mm]{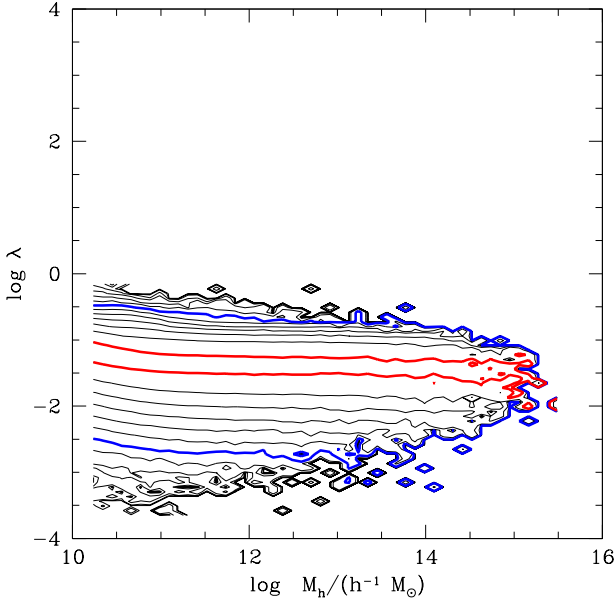}
  \caption{Spin parameter against halo mass for TREE haloes with a
    quasi-equilibrium (QE) limit of 0.5 applied; this can be compared
    with Figs \ref{f:fof_spinmass_all} and \ref{f:tree_spinmass_all}.
    The same contouring is used, i.e. relative halo number density in
    equal logarithmic steps of size $10^{0.5}$.  The inner bold
    contour (red) represents $10^{-1}$ of the haloes at each mass, and
    the outer bold contour (blue) represents $10^{-4}$ of the haloes
    at each mass.  The QE-limit has removed the vast majority of the
    high-spin haloes seen in Fig.  \ref{f:tree_spinmass_all}.}
  \label{f:tree_spinmass_QE}
\end{figure}

The value of $2T/U +1$ for the TREE haloes is plotted against their
mass in Fig. \ref{f:virrat_mass_tree_all}.  We applied a QE cut of
the form 
\begin{equation}
  -Q \leq \frac{2T}{U}+1 \leq Q 
\end{equation}
to the TREEall catalogue, examining the effect of a wide range of
$Q$-values on the $P(\log\lambda)$ and $\lambda(\Mh)$ distributions.
Because of the relatively small numbers of objects with anomalously
high spins, we find that the QE cut makes negligible difference to the
shape of the spin distribution, $P(\log\lambda)$. A very small value,
$Q\la 0.3$, will act to shift the median spin lower by a few percent,
due to the mass dependence seen in Fig \ref{f:virrat_mass_tree_all}.
Through a detailed examination of the $\lambda(\Mh)$ distribution, we
find that a value of $Q$ between $0.4$ and $0.6$ gives a good balance
between removing the mis-defined haloes and reducing the overall
sample size (adding noise and biasing it by mass).  Higher values of
$Q$ allow some mis-defined haloes to creep in, with a significant
impact for $Q\ga 1.0$.  We will use a value of $Q=0.5$ for our cleaned
halo catalogues.  This cut is shown in the horizontal dashed lines of
Fig.  \ref{f:virrat_mass_tree_all}, and the resulting $\lambda(\Mh)$
distribution is shown in Fig. \ref{f:tree_spinmass_QE}.  Applying this
form of virialisation cut on the halo catalogue provides a useful tool
with which to remove haloes with anomalous spins caused by mis-defined
haloes.

\subsubsection{A better halo catalogue: Numerical effects}\label{s:numprobs}
\begin{figure}
  \includegraphics[width=84mm]{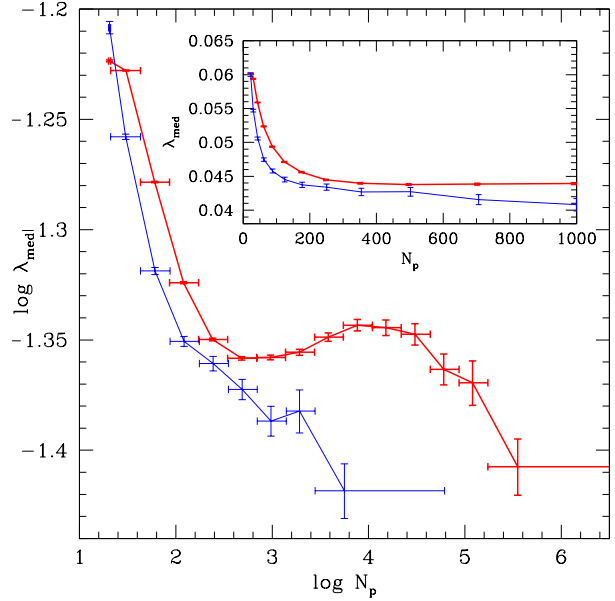}
  \caption{Median spins for all FOF haloes for the Millennium Run
    (thick red line) and the \texttt{milli\_lowres} run (thin blue
    line).  The most massive bin contains 2000 haloes, and the
    remaining bins are logarithmically spaced, with one bin every
    factor of 2 in halo mass (except for the leftmost bin which
    extends down to the cutoff limit of 20 particles.)  Horizontal
    bars mark the bin widths, and vertical error bars give a measure
    of the precision of the median according to Eqn.~\ref{e:mederr}.}
  \label{f:spinmed_np_FOFrestest}
\end{figure}

A second peculiarity of the spin distributions is visible in
Figs. \ref{f:fof_spinmass_all}, \ref{f:tree_spinmass_all} and
\ref{f:tree_spinmass_QE} (for the FOFall, TREEall and
quasi-equilibrium TREE halo catalogues respectively): an upturn in the
spin distribution at low masses.  This can be seen clearer in the
variation of the median spin over mass bins $\lambmed(\Mh)$, plotted
for the FOF haloes in Fig. \ref{f:spinmed_np_FOFrestest}.  This effect
is unrelated to the velocity contamination problems of the mis-defined
haloes, and instead comes from the mass resolution of the simulation
affecting the angular momenta.  This effect has been seen before, for
example by \cite{2005MNRAS.359.1537R} in the context of subhaloes.  To
understand the cause of this effect, consider a continuous object with
angular momentum $\v{J_\rmn{true}}$.  If we construct a realisation of
this object using a sample of $N$ discrete particles, the resulting
angular momentum can be modelled as the vector sum of the `true'
angular momentum (from the continuous object) with a noise vector
oriented in a random direction: $\v{J} = \v{J_\rmn{true}} +
\v{J_\rmn{noise}}$.  This will act to push the measured magnitude $J$
up above $J_\rmn{true}$ because the random direction of
$\v{J_\rmn{noise}}$ will mean it reaches outside the sphere of radius
$|\v{J_\rmn{true}}|$ more than $50\%$ of the time.  Therefore, the
random noise inherent in using discrete particles to sample a
near-continuous object such as a dark matter halo would act to bias
$J$ upwards, and $\lambda$ along with it.

Modern $N$-body codes such as the L-GADGET-II code used for the
Millennium Run are very good at conserving quantities such as energy
and angular momentum, so for a well-resolved object there is
negligible inaccuracy arising from particle discreteness.  For a less
well-resolved object however, the effect can nevertheless be relevant,
even though the angular momentum of the particles making up the halo
has been well conserved.  Discreteness mainly affects the outer parts
of a halo, making the effective surface more jagged than that of the
continuous object it represents.  We also expect the outer parts to
harbour most of the angular momentum.  For lower-mass haloes, a
greater fraction of their particles make up these `surface' regions,
so this problem has a greater effect; the inclusion of a single
particle can make a significant contribution to $\lambda$.  Hence, the
spin parameter rises for haloes with fewer particles because the
discreteness of the haloes' surface layers adds noise to their `true'
angular momenta. (This is not the same effect as discussed in
\cite{2006ApJ...646..815S}; there, a surface pressure term is added to
the virial theorem to account for their halo truncation at $\Rvir$.)

The importance of the noise contribution to $\lambda$ can be examined
by determining the spin distribution of the same simulation (same code
and same corresponding initial condition waves) but ran at a different
resolution.  We performed a lower resolution resimulation of the
Millennium Run, with $2160^3/64=540^3$ particles (so their mass is
$m_\rmn{p,low} = 64 m_\rmn{p,Millen}$), which we will refer to as
\texttt{milli\_lowres}.  The FOF and \textsc{Subfind} algorithms were
implemented on \texttt{milli\_lowres} in the same way as in the
Millennium Run itself. (Although we do not have merger-tree data for
\texttt{milli\_lowres}, and hence cannot construct a TREE halo
catalogue, the Millennium Run results show that the same effect is
seen in FOF, SO and TREE haloes.) Fig. \ref{f:spinmed_np_FOFrestest}
shows the median spin $\lambmed(\Np)$ for FOF haloes in the Millennium
Run and \texttt{milli\_lowres}, with vertical error bars showing a
Gaussian-like estimate of the precision of the median given by:
\begin{equation}
  \epsilon_\mathrm{med}:=
  \frac{(\lambda_{84}-\lambda_{16})}{\sqrt{N_\rmn{halo}}}
  \label{e:mederr}
\end{equation}
where $\lambda_i$ is the $i$th percentile of $\lambda$
($84\%-16\%=68\%$, the amount of data within $\pm 1\sigma$ of a
Gaussian peak) and $N_\rmn{halo}$ is the number of haloes in that mass
bin.  Note that the \emph{spread} in $\lambda$ is much greater than
the uncertainty in $\lambmed$; compare the error bars in Fig.
\ref{f:spinmed_np_FOFrestest} with the data shown in Fig.
\ref{f:fof_spinmass_all}.

Although the two curves show the same qualitative behaviour (the
low-$\Np$ upturn), there is a vertical shift between them.  This is
due to \texttt{milli\_lowres} containing fewer of the mis-defined
objects described in Section~\ref{s:groupprobs} than the Millennium
Run itself.  Decreasing the resolution effectively smooths the density
field, so that small objects with more massive neighbours can
disappear completely, whereas a more isolated object of the same mass
may still survive (although containing fewer particles).  This means
that the ``real'' objects are retained (and there are still many
under-resolved ones causing the upturn in $\lambmed(\Np)$), but there
is a reduction in the number of mis-defined objects. The
\texttt{milli\_lowres} results seem to confirm the dominance of
numerical effects at low-$\Np$, above which the effect of noise is
negligible.

\begin{figure}
  \includegraphics[width=84mm]{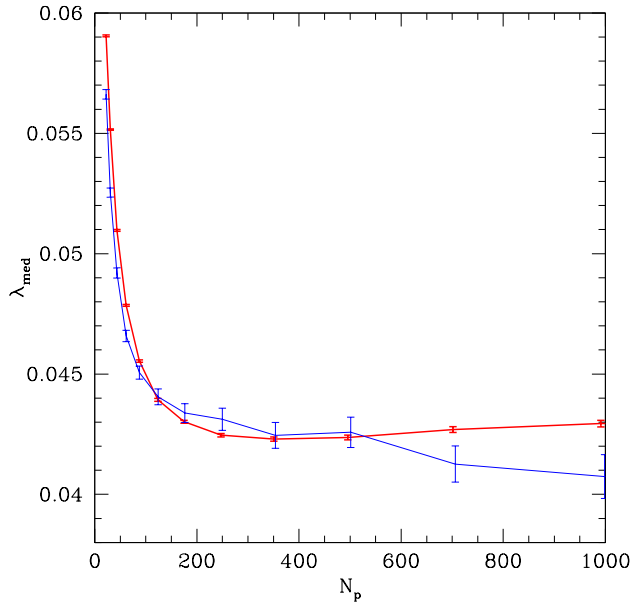}
  \caption{Median spins against number of particles in haloes, for
    FOF haloes with the QE cut applied.  The thin blue line shows the
    \texttt{milli\_lowres} haloes and the thick red line shows those
    from the Millennium Run itself.  The error bars show the uncertainty
    on the median, using Eqn.~\ref{e:mederr}.  The two lines show an
    identical trend at low $\Np$, demonstrating that the upturn in
    $\lambmed$ is indeed a numerical effect, affecting the spins of
    haloes containing fewer than about $300$ particles.}
  \label{f:spin_np_FOFrestest_QE}
\end{figure}

If we apply the quasi-equilibrium cut described in section
\ref{s:removal} to remove the mis-defined objects, we can examine the
effect of discreteness on the median spins of just the `real' haloes.
Fig.~\ref{f:spin_np_FOFrestest_QE} shows $\lambmed(\Np)$ for
QE-selected FOF haloes, compared with QE-selected FOF haloes from
\texttt{milli\_lowres}.  In contrast to
Fig.~\ref{f:spinmed_np_FOFrestest}, the two lines now lie on top of
each other, exhibiting the same upturn in spin for haloes with the
same number of particles. This demonstrates that the upturn is purely
a numerical effect. We can exclude haloes that appear to be dominated
by this effect by fixing a limit of $\Np\ge 300$ on the halo
catalogue.

Fig. \ref{f:virrat_mass_tree_all} shows how the QE and $\Np$ cuts we
use relate to one another for the TREE haloes.  The low-$\Np$ cut has
a significantly stronger effect on the halo catalogue.  Applying the
QE cut on its own reduces the population of the TREEall catalogue by
only $0.3\%$; applying the low-$\Np$ limit as well means removing in
total $91.17\%$ of the original TREEall haloes (see Table
\ref{t:halocats}).  We refer to the resulting cleaned TREE-halo
catalogue as TREEclean, and it is these haloes whose properties we
shall be examining in detail.  In some cases, for completeness, we
shall compare the results with those from the FOFclean and SOclean
catalogues.  These are cleaned using the same $Q=0.5$ and $\Np\ge 300$
cuts as the TREEclean catalogue.

It is important to note that the criteria we have adopted -- the QE
cut and low-$\Np$ limit -- are those appropriate to the quantities of
interest in this work.  For example, Fausti-Neto et al (in prep.) use
the substructure parameter $S = \Delta r/\Rvir$, where $\Delta r$ is
the distance between the potential minimum of the halo and its centre
of mass.  Their final criteria for selecting haloes in the Millennium
simulation are $S<0.07$ and $\frac{2T}{U}+1 > -0.25$.  They are
concerned with fitting density profiles to haloes, and using the
substructure parameter allows them to remove haloes with a large
fraction of mass in substructures which would otherwise contaminate
their results.  Similarly, \cite{2006astro.ph..8157M} define an
``offset parameter'' as $x_\rmn{off}=\Delta r/\Rvir$, where $\Delta r$
is measured from the most-bound particle rather than the potential
minimum.  They use this alongside the rms of their density profile
fits, to assess the quality of their halo catalogues for estimates of
halo concentration.  Although we have examined the substructure/offset
parameter and how it affects $\lambda$, we find that it is not useful
in removing the mis-defined haloes, or those whose spins are dominated
by the numerical effects discussed above.

Having successfully implemented an appropriate groupfinder and cleaned
the resulting halo catalogues, we can now proceed to examine their
spin properties.  The FOF and TREE halo catalogues, including some
halo properties and semi-analytic galaxy properties, are publicly
available online\footnote{http://www.mpa-garching.mpg.de/millennium/}
\citep{2006astro.ph..8019L}.


\section{Results}\label{s:results}
\subsection{The form of the spin distribution}
\begin{figure}
  \includegraphics[width=84mm]{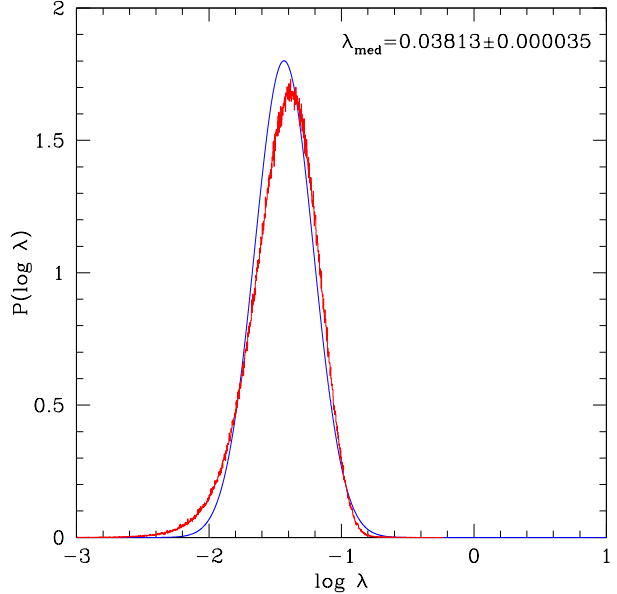}
  \caption{The normalised spin distribution (red histogram) for
    merger-tree haloes with $\Np>300$ and filtered using the QE
    criterion (the TREEclean catalogue).  The best-fitting single
    Gaussian function is plotted as the smooth blue curve.  The
    Gaussian drops too quickly at low spins, and too slowly at higher
    spins.  To try to minimise these effects, the best fit has a peak
    location that is shifted away from that of the histogram data.
    The median spin of the distribution is also displayed.}
  \label{f:spinhist_Tree_QE_Np_1gauss}
\end{figure}

The median spin of the TREEclean halo catalogue is $\lambmed=0.0381$.
The distribution of halo spins about the median, $P(\lambda)$, has
been often fitted with a lognormal function (e.g.
\citealt{1998ApJ...507..601V, 2001ApJ...557..616G,
2005ApJ...627..647B}), i.e. a Gaussian in $\log\lambda$:
\begin{equation}
  P(\log\lambda) = \frac{1}{\siglg\sqrt{2\pi}}
                   \exp\left[
		     -\frac{\log^2\left(\lambda/\lambda_0\right)}
                           {2\siglg^2}
		   \right]
  \label{e:gaussian}
\end{equation}
While this fitting function has proved adequate for small numbers of
objects, we find that for the $>10^6$ haloes in the Millennium
simulation, deviations from a Gaussian are clear and significant.  The
spin distribution drops faster than a Gaussian at high spins, and
slower than a Gaussian at low spins.  The best fit to the TREEclean
catalogue is shown in Fig. \ref{f:spinhist_Tree_QE_Np_1gauss}, which
fits Eqn.~\ref{e:gaussian} with peak location $\lambda_0=0.03687\pm
0.000016$ and width $\siglg=0.2216\pm 0.00012$.\footnote{Throughout
this paper, the quoted uncertainties on best-fitting parameters are
given by the square root of the diagonal of the covariance matrix for
that fit.}  The corresponding lognormal function of $\lambda$ has the
same peak, and a width of $\sigma=\ln(10)\siglg$.  The fit has a
reduced-$\chi^2$ of $40.46$.

Part of the reason why a lognormal is such a poor fit is that this
function strongly avoids very low spin values, whereas the real
distribution, based as it is on the three-dimensional vector $\v{j}$,
does not.  The longer tail at low-$\lambda$ is primarily due to the
distribution of $\v{j}$ being smooth and isotropic about $j=0$,
implying\footnote{ \cite{2006astro.ph..8157M} claimed the
low-$\lambda$ tail is due to the higher uncertainty in $\lambda$ at
low values.  However, by varying the minimum $\Np$ for the halo
catalogue, and hence the uncertainty in $\lambda$, we found that the
low-$\lambda$ side of the distribution consistently drops off slower
than the high-$\lambda$ end, confirming that this shape is not
primarily due to uncertainties.} that $P(\log\lambda)\propto
\lambda^3$.

\begin{figure}
  \includegraphics[width=84mm]{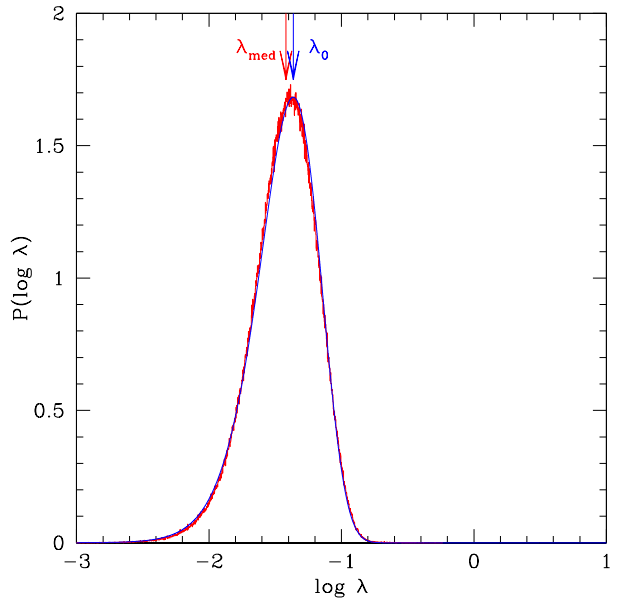}
  \caption{Histogram (red) of the normalised $\log\lambda$
    distribution of TREEclean haloes, as in Fig.
    \ref{f:spinhist_Tree_QE_Np_1gauss}. The smooth solid curve (blue)
    is the best fit to the data using Eqn~\ref{e:arjfn}. The peak
    location of the fit ($\lambda_0$) and the median of the data
    ($\lambmed$) are marked with arrows.}
  \label{f:spinhist_Tree_QE_Np_arj}
\end{figure}

We have found that the following function provides a better
description of the data:
\begin{equation}
  P(\log\lambda) = A \left(\frac{\lambda}{\lambda_0}\right)^3
         \exp\left[-\alpha
         \left(\frac{\lambda}{\lambda_0}\right)^{3/\alpha}
	      \right]
  \label{e:arjfn}
\end{equation}
For the normalised spin distribution, we can express $A$ in terms of
the other free parameters, $\alpha$ and $\lambda_0$ (the peak
location):
\begin{equation}
  A = 3\ln10\frac{\alpha^{\alpha-1}}{\Gamma(\alpha)}
\end{equation}
where the gamma function $\Gamma(\alpha)=(\alpha-1)!$.  The best fit
to the data is shown in Fig.  \ref{f:spinhist_Tree_QE_Np_arj}, and
has parameters:
\begin{eqnarray*}
  \lambda_0 = 0.04326\pm 0.000020 & & \alpha=2.509\pm 0.0033
\end{eqnarray*}
with a much-improved reduced-$\chi^2$ of $2.58$.

We have examined whether the deviations from lognormal depends on our
choice of the quasi-equilibrium parameter $Q$ when cleaning the halo
catalogue.  We found that the fit remains good over a wide range of
$Q$.  We have also found the best fit of Eqn. \ref{e:arjfn} to the
FOFclean and SOclean catalogues.  The results for SOclean haloes are
remarkably similar to those for the TREEclean haloes, with a
reduced-$\chi^2$ of $3.10$:
\begin{eqnarray*}
  \lambda_0 = 0.04174\pm 0.000022 & & \alpha=2.540\pm 0.0036
\end{eqnarray*}
The median spin of the distribution is $\lambmed=0.0367$.  The haloes
in the FOFclean catalogue, which have a median spin of
$\lambmed=0.04288$, are not as well fitted by Eqn.\ref{e:arjfn}.  The
reduced-$\chi^2$ is $15.0$ and the parameter values are:
\begin{eqnarray*}
  \lambda_0 = 0.04929\pm 0.000027 & & \alpha=3.220\pm 0.0046
\end{eqnarray*}
This is, in fact, slightly worse than the best-fitting lognormal (Eqn.
\ref{e:gaussian}), which yields a reduced-$\chi^2$ of $12.5$, with a peak
location $\lambda_0=0.04222\pm 0.000022$ and width $\siglg=0.2611\pm
0.00016$.

These tests show that the distribution of $\lambda$ depends on the
careful definition and selection of dark matter haloes.  However, the
fact that the distribution is non-lognormal is not a consequence of
the particular choice of groupfinder or selection criteria used here
-- the form of the distribution is not peculiar to the TREEclean
catalogue.

\subsection{Correlation of spin and halo mass}
\begin{figure}
  \includegraphics[width=84mm]{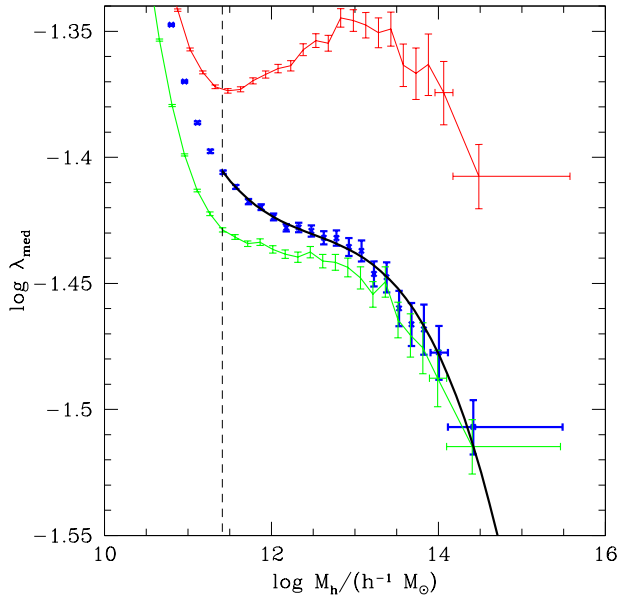}
  \caption{The median spins of halo mass bins, for TREEclean haloes
    (bold blue points), SOclean haloes (medium green line) and
    FOFclean haloes (thin red line).  Data have also been calculated
    below the $300$-particle low-mass limit (marked with a dashed
    line).  The thick black curve is the best-fitting cubic polynomial
    to the TREEclean data.  The vertical error bars are given by
    Eqn~\ref{e:mederr}.  The mass-binning scheme is similar to Fig.
    \ref{f:spinmed_np_FOFrestest}, but with bins every factor of
    $\sqrt{2}$ in mass.  Only the widths of the most massive two bins
    are marked, for clarity.}
  \label{f:spinmed_QEnp_comp}
\end{figure}

The variation of median spin parameter with halo mass, for the cleaned
catalogues from the three groupfinders is shown in
Fig. \ref{f:spinmed_QEnp_comp}.  It is interesting to note that the
FOF haloes exhibit an upturn in spin for objects more massive than the
low-$\Np$ cut, an effect that is not present in the TREE or SO haloes.
This can be attributed to the outer parts of the FOF haloes consisting
mainly of unbound `fuzz' particles.  These will usually have higher
velocities, which act to inflate the spin.  These particles are not
part of the TREE haloes, and most will be shaved off in SO haloes too.
The SO and TREE haloes show a shallow downwards trend of $\lambda_{\rm
med}$ up to $\Mh \sim 10^{13}\munit \sim 12000\mp$ and a rapid decline
at larger masses.

We fit a cubic polynomial to the TREEclean median spin data,
\begin{equation}\label{e:poly}
  \log\lambmed = \alpha x^3 + \beta x^2 + \gamma x + \delta
\end{equation}
where $x=\log\Mh/(\munit)$.  The best-fitting values of these parameters
are:
\begin{eqnarray*}
  \alpha &=& (-8.6 \pm 1.4  )\times 10^{-3}  \\
  \beta  &=& ( 3.2 \pm 0.54 )\times 10^{-1} \\
  \gamma &=&  -4.1 \pm 0.68 \\
  \delta &=&  15.7  \pm 2.8
\end{eqnarray*}
with a reduced-$\chi^2$ of $0.44$.

While the trend of $\lambda_{\rm med}$ with mass is real, it is
important to note that it is a small effect; the scatter around this
median is large (compare with Fig. \ref{f:tree_spinmass_QE}, which
shows data from the same haloes but on a $\log\lambda$ scale).
This is in qualitative agreement with previous results
(e.g. \citealt{1996MNRAS.281..716C}), but because of its weak nature,
this trend has often not been visible
(e.g. \citealt{1992ApJ...399..405W, 1999MNRAS.302..111L,
2006astro.ph..8157M}).

\subsection{The halo shape distribution}
\begin{figure}
  \includegraphics[width=84mm]{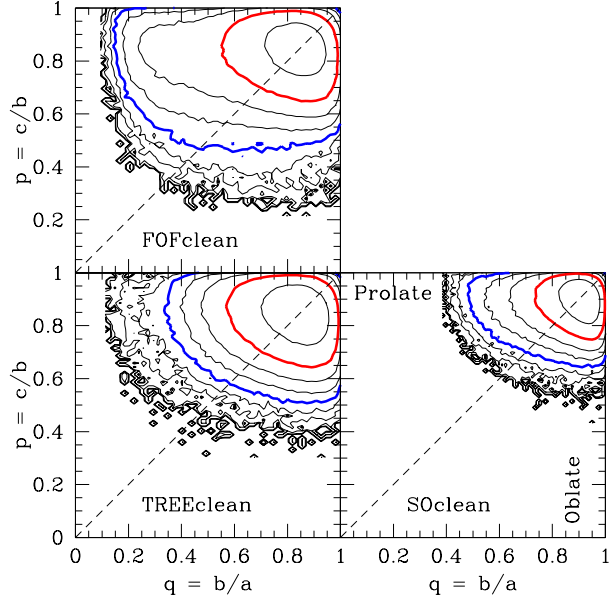}
  \caption{Axis ratios $p$ and $q$ for the cleaned TREE, FOF and SO
    halo catalogues.  Prolate objects have $p=1$,
    oblate objects have $q=1$, and spherical objects have $p=q=1$.
    The SO haloes are more spherical than the other two types.  TREE
    and FOF objects exhibit a range of shapes, and all three
    catalogues show a preference for prolateness over oblateness.  The
    contouring is by halo numbers, in equal logarithmic steps of
    $10^{0.5}$.  In each plot, the red inner bold contour represents
    $10^{3}$ haloes, and the blue outer bold contour represents
    $10^{1.5}$ haloes.  }
  \label{f:shapeshape}
\end{figure}

The shapes of the haloes are described by the axes, $a\ge b\ge c$, of
the ellipsoid derived from the inertia tensor, as described in
Section~\ref{s:properties}.  Fig. \ref{f:shapeshape} shows the
distributions of $p=c/b$ and $q=b/a$ for the three cleaned halo
catalogues.  The haloes are generally triaxial, but they have a range
of shapes, with a slight preference for prolateness over oblateness.
The distribution agrees qualitatively with previous work such as that
by \cite{1988ApJ...327..507F}, \cite{1992ApJ...399..405W},
\cite{1996MNRAS.281..716C}, \cite{2002A&A...395....1F} and
\cite{2005ApJ...627..647B}.  Unsurprisingly, SO haloes are more
spherical than FOF or TREE haloes.  FOF haloes show a much broader
distribution of shapes (and a stronger preference for prolateness)
than SO or TREE haloes.

\begin{figure}
  \includegraphics[width=84mm]{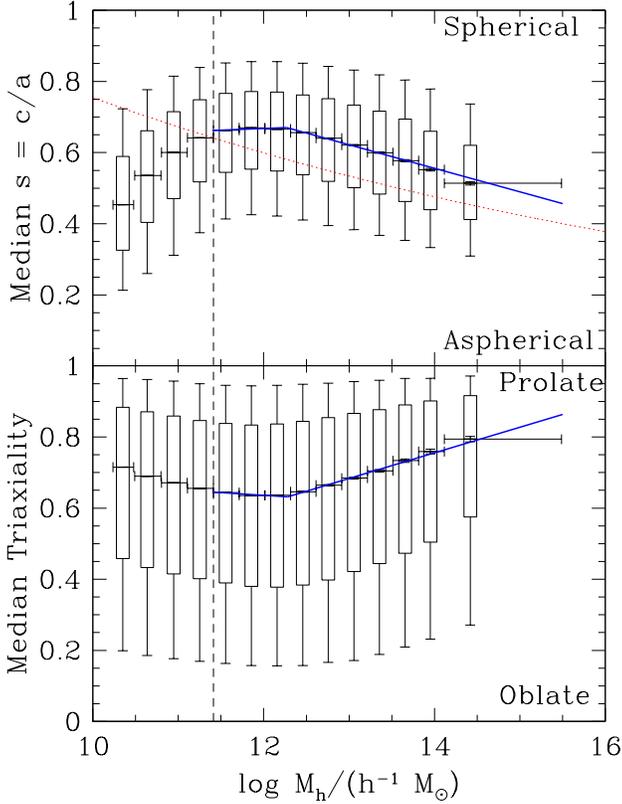}
  \caption{The median, in bins of halo mass, of the axis ratio $s$
    (top) and the triaxiality parameter $\mathcal{T}$ (bottom), for
    TREEclean haloes, using the same mass binning scheme as in Fig.
    \ref{f:spinmed_np_FOFrestest}.  The low-$\Np$ limit of 300
    particles is marked with a dashed line, and data points have also been
    plotted below this limit).  The medians for FOFclean and SOclean
    haloes follow the same behaviour as the TREEclean haloes, but with
    SO objects being more spherical and FOF objects less; FOF haloes
    are more prolate and SO haloes show a weaker preference for
    prolateness.  Error bars on the medians (following
    Eqn.~\ref{e:mederr}) are plotted, but most are vanishingly small.
    The whisker bars are percentiles, at the equivalent of $1\sigma$
    ($68\%$ of haloes, boxes) and $2\sigma$ ($95\%$ of haloes, bars);
    from the bottom to the top of each graph, these show where
    approximately $2.5\%$, $16\%$, $84\%$, and $97.5\%$ of the haloes
    have $s$ (or $\mathcal{T}$) below these values.  The thick blue
    lines show broken-line fits to the data; see the text for details.
    The dotted red line is the fit of \citealt{2006MNRAS.367.1781A},
    who used a different definition of a halo (see text).}
  \label{f:spheric_triax_mass}
\end{figure}

Fig.~\ref{f:spheric_triax_mass} shows how the median shape of haloes
in the TREEclean catalogue changes with halo mass, using the
minor-to-major axis ratio $s$ and the triaxiality parameter
$\mathcal{T} = (a^2-b^2)/(a^2-c^2)$.  More massive haloes tend be less
spherical and more prolate.  Again, this is in qualitative agreement
with previous results, such as those of \cite{1992ApJ...399..405W},
\cite{2002sgdh.conf..109B}, \cite{2005ApJ...629..781K},
\cite{2006ApJ...646..815S}, \cite{2006EAS....20...25G},
\cite{2006MNRAS.367.1781A}, and \cite{2006astro.ph..8157M}.  This is
also what one might expect in a hierarchical formation model in which
haloes tend to form by matter collapsing along filaments, leading to
prolateness, rather than onto sheets which would lead to oblateness.
Furthermore, the more massive haloes form later, and have had less
time to relax into more spherical configurations.  Since we have
deliberately tried to select the more relaxed objects, the remaining
trend we see here is weak.  Furthermore, although the medians follow
well-defined trends, the spread of the distribution in halo shapes
covers virtually the entire range in both $s$ and $\mathcal{T}$, as
can be seen from the percentile bars on the graph.

The two graphs in Fig. \ref{f:spheric_triax_mass} both show a change
in behaviour not seen in previous work, around the $\Np=300$ limit.
Resolution tests similar to those described in section
\ref{s:numprobs} were carried out using the \texttt{milli\_lowres}
simulation, to assess whether this change in behaviour was indeed a
numerical effect similar to that seen in halo spins (in
Figs. \ref{f:spinmed_np_FOFrestest}, \ref{f:spin_np_FOFrestest_QE} and
\ref{f:spinmed_QEnp_comp}).  The results showed that these halo shape
parameters do also require $\Np\ga 300$, reinforcing our previous
choice.  Indeed, one would expect haloes whose spins are affected by
particle discreteness (i.e. with $\Np\la 300$) to be less spherical
and more stringy (prolate).

We fit a broken line to both shape parameters for the TREEclean
catalogue, of the form:
\begin{equation}\label{e:broken}
  y_\mathrm{med}(x) = \left\{
    \begin{array}{ll}
      m_1 x + c_1   &  x\le x_0 \\
      m_2 x + c_2   &  x\ge x_0 \\
    \end{array}
    \right.
\end{equation}
where $x=\log\Mh/(\munit)$.  We fit with $m_1$, $m_2$, $x_0$ and $c_2$
as free parameters, with $c_1 \equiv c_2 + (m_2-m_1)x_0$.  The fitted
parameters for $s_\mathrm{med}$ are:
\begin{eqnarray*}
  m_{1,s} = (9.2\pm 0.87)\times 10^{-3},
      & & c_{1,s} = 0.56 \pm 0.015, \\
  m_{2,s} = (-6.6\pm 0.12)\times 10^{-2},
      & & c_{2,s} = 1.48\pm 0.015, \\
  x_{0,s} = 12.27\pm 0.012 & &
\end{eqnarray*}
with a reduced-$\chi^2$ of $29.9$.  The fitted parameters for
$\mathcal{T}_\mathrm{med}$ are:
\begin{eqnarray*}
  m_{1,\mathcal{T}} = (-1.6\pm 0.18)\times 10^{-2},
      & & c_{1,\mathcal{T}} = 0.82 \pm 0.031, \\
  m_{2,\mathcal{T}} = (7.2\pm 0.24)\times 10^{-2},
      & & c_{2,\mathcal{T}} = -0.25 \pm 0.029, \\
  x_{0,\mathcal{T}} = 12.28\pm 0.021 & &
\end{eqnarray*}
with a reduced-$\chi^2$ of $4.27$.  The two mass breakpoints $x_{0,s}$
and $x_{0,\mathcal{T}}$ agree within their uncertainties.  

\cite{2006MNRAS.367.1781A} fit a power-law to $s_\mathrm{med}(\Mh)$.
This is plotted in Fig. \ref{f:spheric_triax_mass}, and indicates that
that their haloes are significantly less spherical than ours.  This is
largely a result of different group definitions; although not plotted,
we find that our SOclean and FOFclean catalogues differ from the
TREEclean results by a similar amount.  A power-law of the type used
by \cite{2006MNRAS.367.1781A} would not be a good fit to the data
presented here which have a definite change in slope towards lower
halo masses.

The overall distributions of $s$ and $\mathcal{T}$ are shown in the
upper plots of Figs. \ref{f:spheric_spin} and \ref{f:triax_spin}.
These agree well with distributions seen previous work,
e.g. \cite{2005ApJ...627..647B} and \cite{2006ApJ...646..815S}.  The
drop-off in halo sphericity below about $s\sim0.3$ can be explained by
considering how flatter haloes would puff up due to bending
instabilities \citep{1994ApJ...425..551M}.

\subsection{Spin and shape parameters}
\begin{figure}
  \includegraphics[width=84mm]{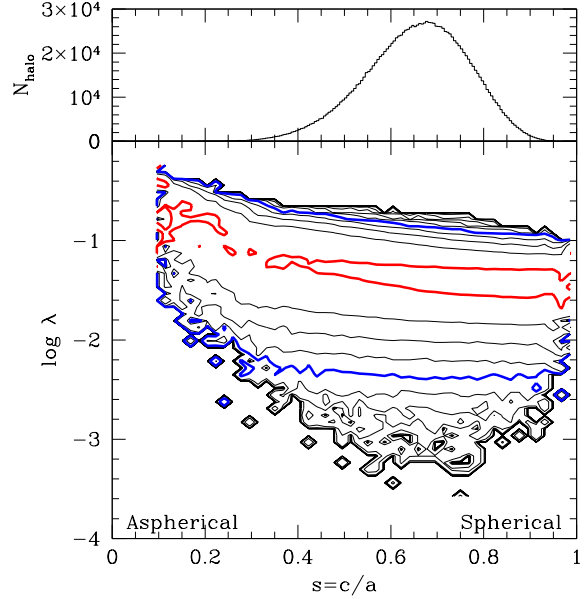}
  \caption{The main plot shows spin versus axis ratio $s=c/a$ for
    TREEclean haloes; contouring is similar to Fig.
    \ref{f:fof_spinmass_all}, showing the number density of haloes,
    normalised by the number of haloes in each $s$-bin.  The red inner
    bold contour represents $10^{-1}$ of the haloes in each $s$-bin,
    the blue outer bold contour represents $10^{-3}$ of the haloes in
    each $s$-bin, and the contours are spaced in equal logarithmic
    steps of $10^{0.5}$.  The upper plot is a histogram of $s$ for
    TREEclean haloes, effectively showing the function by which
    the contour plot has been normalised.}
  \label{f:spheric_spin}
\end{figure}

\begin{figure}
  \includegraphics[width=84mm]{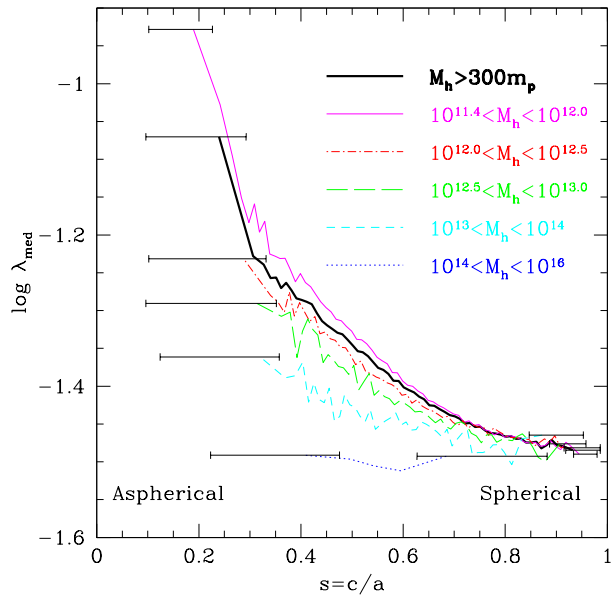}
  \caption{The median spin parameter in bins of axis ratio $s=c/a$,
    for haloes from the TREEclean catalogue.  The heavy black line is
    for all haloes with more than $300$ particles (the full TREEclean
    sample); the other lines show the trends for different halo mass
    bins.  The horizontal bars show the widths of the first and last
    bins for each line.  The trend is for more spherical haloes to
    have less coherent rotation in the median.  This trend becomes
    very steep for the most aspherical haloes (although these are not
    present at higher masses because of the rapid drop in the halo
    mass function).}
  \label{f:spheric_spinmed} 
\end{figure}

\begin{figure}
  \includegraphics[width=84mm]{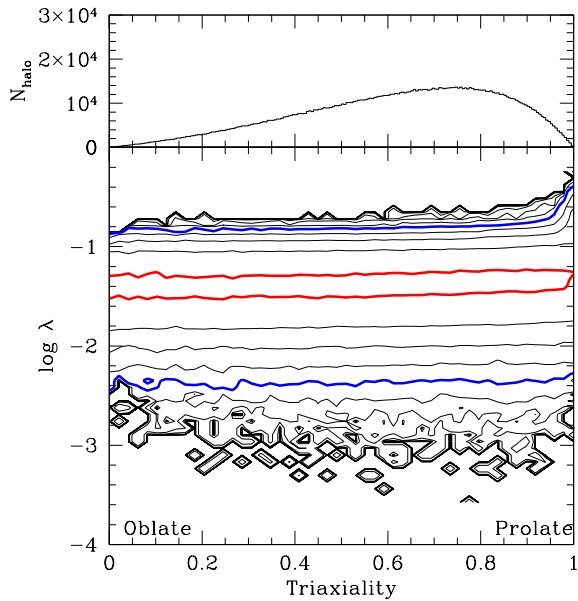}
  \caption{The main plot shows spin versus triaxiality parameter
    $\mathcal{T}$ for TREEclean haloes; contouring is as for Fig.
    \ref{f:spheric_spin}, i.e. in equal logarithmic steps of
    $10^{0.5}$ haloes per $\mathcal{T}$-bin.  The red inner bold
    contour represents $10^{-1}$ of the haloes in each
    $\mathcal{T}$-bin, and the blue outer bold contour represents
    $10^{-3}$ of the haloes in each $\mathcal{T}$-bin.  The upper plot
    is a histogram of $\mathcal{T}$ for TREEclean haloes,
    effectively showing the function by which the contour plot has
    been normalised.}
  \label{f:triax_spin}
\end{figure}

The  relationship  between  spin  parameter  and  halo  shape  is
illustrated  in Figs  \ref{f:spheric_spin}, \ref{f:spheric_spinmed}
and \ref{f:triax_spin}.
Fig. \ref{f:spheric_spinmed} emphasises the trend visible in
Fig. \ref{f:spheric_spin} by plotting the median spin parameter as a
function of $s$ for different mass bins.  There is a clear trend for
more spherical haloes to exhibit less coherent rotation.  Although
this trend is in the sense one might na\"{\i}vely expect, the haloes,
in fact, do not have very high spin, and are not rotationally
supported.  The origin of this trend is likely to lie instead in the 
effects of the tidal torques experienced by the haloes during their
early phases of formation.

As seen previously in Fig.~\ref{f:spinmed_QEnp_comp}, the least
massive objects have the most extreme spins, in the median.
Fig.~\ref{f:spheric_spinmed} shows that the higher spin objects are
also less spherical.  The haloes which are closest to spherical
have a spin parameter that is independent of halo mass, and has
$\lambmed\approx 0.033$. (This does not apply to the most massive
haloes, however, since their population lacks the more spherical
objects)  Furthermore, the median spins for the more massive haloes
are independent of shape and have $\lambmed\approx 0.032$.

In contrast to the variation of $\lambmed$ with $s$,
Fig. \ref{f:triax_spin} shows that there is only a very weak trend of
spin with halo triaxiality. Over the entire range of triaxiality, each
$\mathcal{T}$-bin contains a very similar fraction of haloes at each
value of $\log\lambda$.

\subsection{Spin-shape alignment}
\begin{figure}
  \includegraphics[width=84mm]{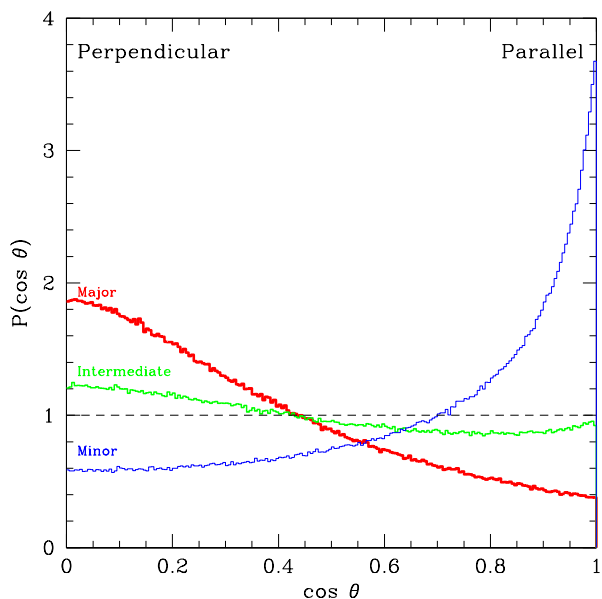}
  \caption{Normalised histograms of the cosines of the angle between
    the angular momentum vector and the major (thick red line),
    intermediate (medium green line) and minor (thin blue line) axes
    of haloes in the TREEclean catalogue, as defined in the text.
    A random distribution would be a flat line at $P(\cos\theta)=1$. }
  \label{f:spin_shape_align}
\end{figure}

Fig.~\ref{f:spin_shape_align} shows the angle between the angular
momentum vector and the three shape axis vectors, e.g.:
\begin{equation}
  \cos\theta_a  = \left| \v{j}\dotprod\v{\hat{a}}/j\right|
\end{equation}  
for alignment with the semimajor axis given by the unit vector
$\v{\hat{a}}$.  Note that this definition does not distinguish between
$\v{j}$ lying parallel or antiparallel to the axis vectors.

Most haloes have their spin axis well aligned with their minor axis,
and lying perpendicular to their major axis.  However, the
distribution of alignments with respect to all three axes is fairly
broad.  This agrees with previous results 
(e.g. \citealt{1992ApJ...399..405W}, \citealt{2005ApJ...627..647B},
\citealt{2006MNRAS.367.1781A} and \citealt{2006ApJ...646..815S}).

\begin{figure}
  \includegraphics[width=84mm]{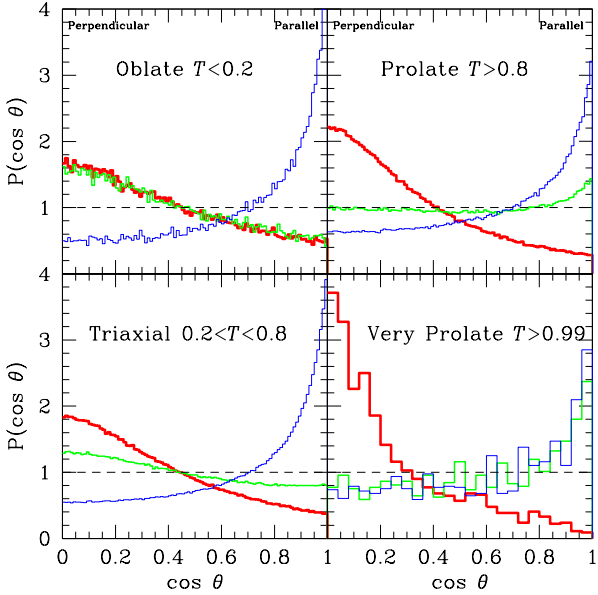}
  \caption{Normalised histograms of the cosine of the angle between
    the specific angular momentum vector and the major (thick red
    line), intermediate (medium green line) and minor (thin blue line)
    axes of the haloes.  The TREEclean catalogue has been cut at the
    values of the triaxiality parameter shown.  The most spherical
    haloes ($s>0.8$) have been removed from each sample.  The `very
    prolate' selection only contains $1360$ groups, giving rise to a
    noisier histogram.}
  \label{f:spin_shape_alignTcuts}
\end{figure}

Extremely oblate objects have a degeneracy between the major and
intermediate axes ($\frac{b}{a}\sim 1$), so there is an equal
probability for the angular momentum vector to subtend a given angle
with either axis This can be seen in the top-left panel of
Fig.~\ref{f:spin_shape_alignTcuts}, which shows the alignment
distribution for the more oblate haloes, i.e. those with
$\mathcal{T}<0.2$ (and $s<0.8$ in order to avoid an $a\sim b\sim c$
degeneracy).  Since most haloes have $\v{j}$ aligned with
$\v{\hat{c}}$, $\v{j}$ has a preference for being at right angles to
the two large axes.

For extremely prolate haloes, the degeneracy is between minor and
intermediate axes ($\frac{c}{b}\sim 1$). In this case, the
distribution of the alignment of the angular momentum vector with
either of these axes is similar only for the tail of extremely prolate
haloes, $\mathcal{T}>0.99$ (bottom-right panel). For $\mathcal{T}>0.8$,
only a small number of haloes have their spin axis aligned with the
intermediate axis (top-right panel); like the bulk of the population
as a whole, most of these haloes rotate around
their smallest axis.

The reason for the distinction between the major and intermediate axes
present in Fig.~\ref{f:spin_shape_align} can now be seen to be a
combination of various effects.  The preference for prolateness over
oblateness means that there is a tendency for the intermediate axis to
be more similar to the minor than to the major axis. This increases
the probability of $\v{j}$ being aligned to the $\v{\hat{b}}$ axis.
However, since $\v{j}$ can be aligned with only one axis (and when this
happens it must be perpendicular to the other two), the preference for the
spin axis to be the minor axis outweighs the preference for
prolateness, and this results in the slight excess probability for $\v{j}$
to be perpendicular to the intermediate axis seen in the figure.

\subsection{Spin, shape and Halo clustering}
\begin{figure*}  
  \includegraphics[width=150mm]{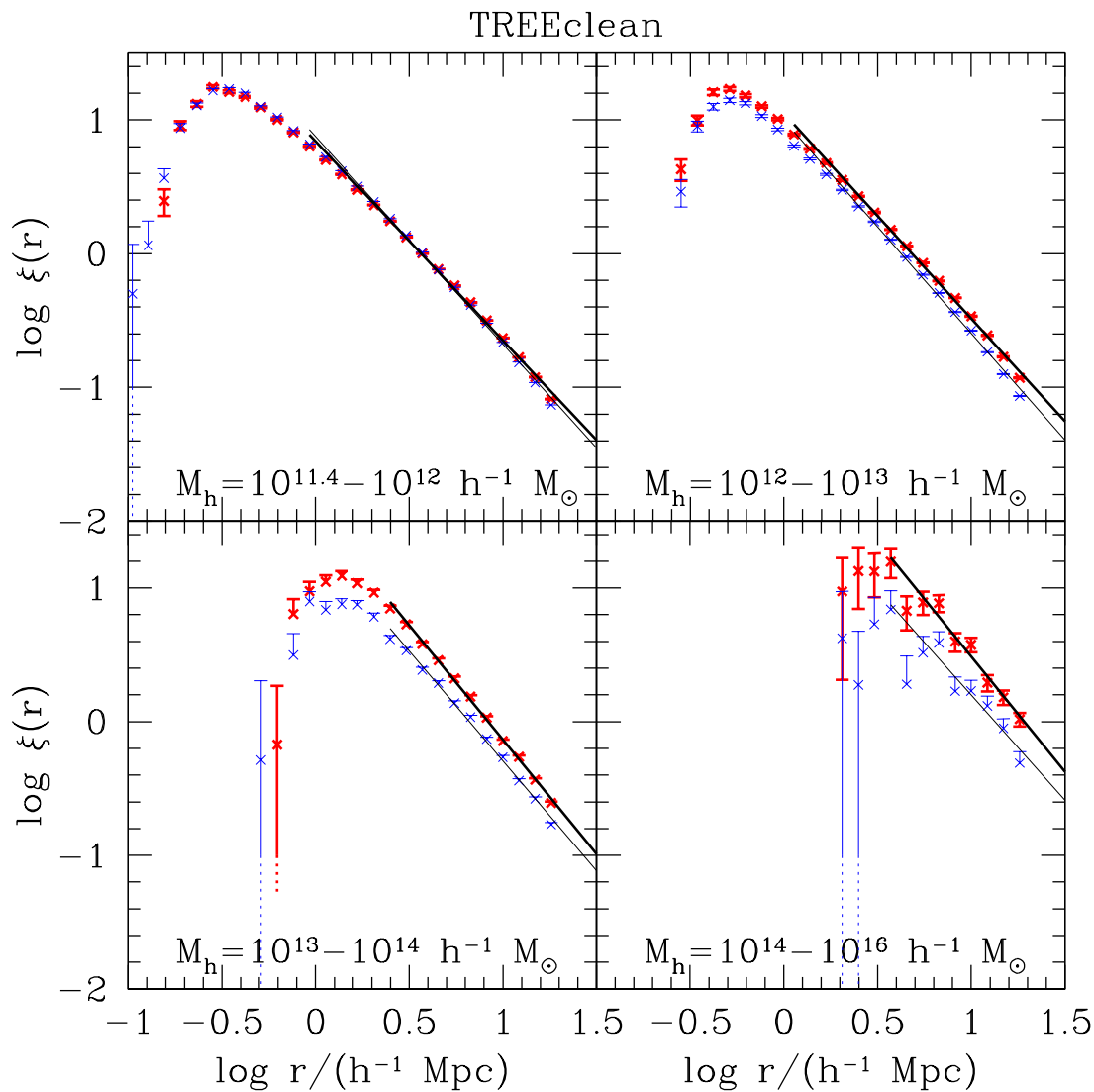}
  \caption{Two-point correlation function, $\xi(r)$, for TREEclean
    haloes in four mass bins.  The lines represent power-law fits over
    the $r$-bins shown.  The thick lines and red points are for haloes
    with $\lambda\geq\lambmed$ for that mass bin; the thin lines and
    blue points are for haloes with $\lambda<\lambmed$.  The data are
    noisy in the higher mass bins which contain fewer haloes. The
    error bars are Poisson errors, i.e.  the square-root of the number
    of pairs in each $r$-bin, divided by the mean number of pairs per
    $r$-bin.}
  \label{f:xir_QE_np_TREE}
\end{figure*}

\begin{figure*}
  \includegraphics[width=150mm]{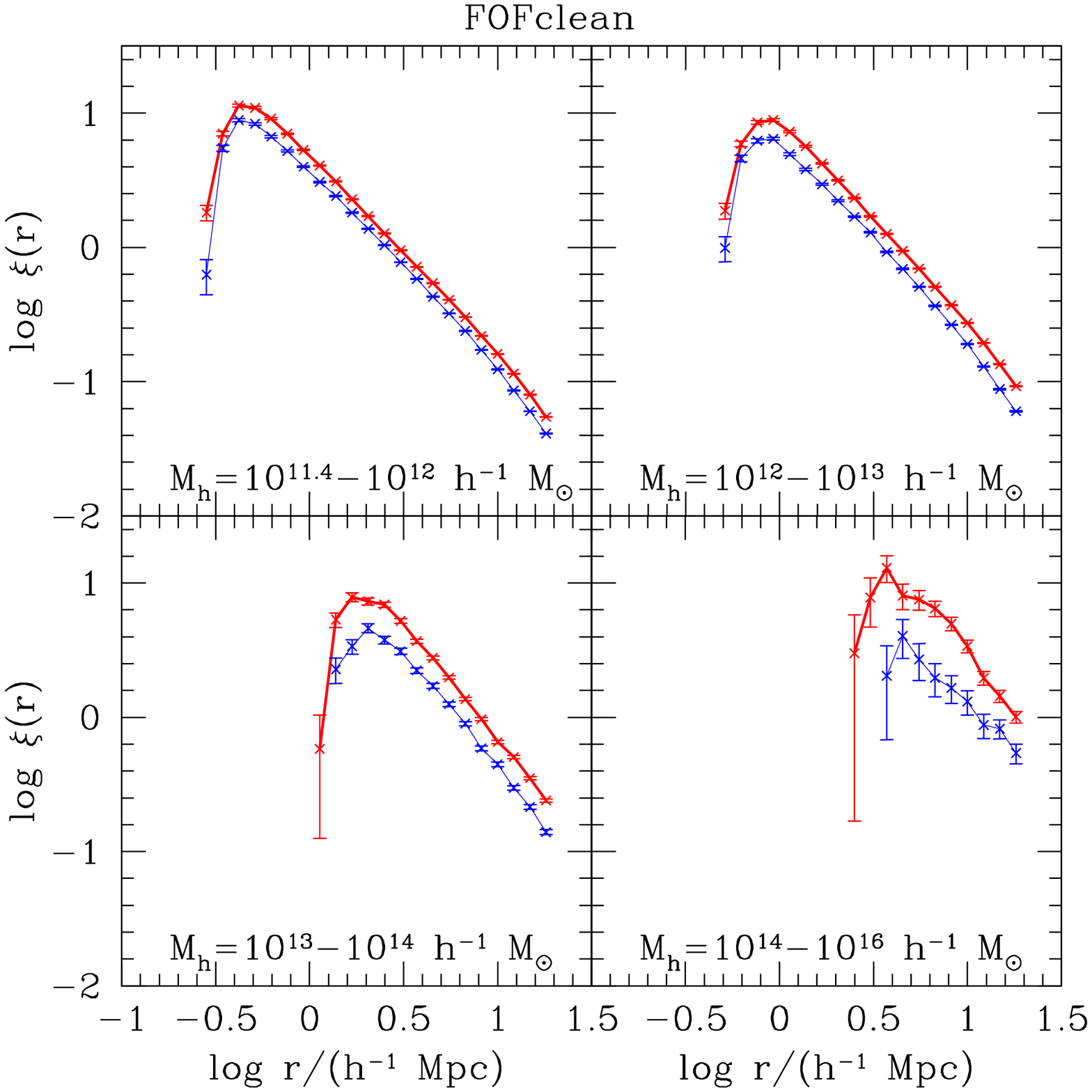}
  \caption{As Fig. \ref{f:xir_QE_np_TREE}, but for the FOFclean
    haloes.  No fits were made to these data, and the lines merely
    join the points.  For this catalogue, the higher spin haloes
    (thick red lines) are consistently and significantly more strongly
    clustered that the lower spin haloes (thin blue lines).  The results
    for the SOclean haloes are very similar.}
  \label{f:xir_QE_np_FOF}
\end{figure*}

In this section, we investigate whether halo spin and shape has an
environmental dependence. We quantify environment by means of the
haloes' two-point correlation function, $\xi(r)$, and we explore
whether the clustering amplitude differs for haloes with different
spin.  The three halo catalogues were divided into four bins in mass,
and the haloes in each mass bin were then divided about the median
spin for that mass; $\xi(r)$ was calculated for each set of haloes.
The results are plotted in Figs~\ref{f:xir_QE_np_TREE}
and~\ref{f:xir_QE_np_FOF} for the TREEclean and FOFclean
catalogues respectively. (The results for SOclean haloes are very
similar to the FOF results.)  The lines in Fig.~\ref{f:xir_QE_np_TREE}
show power-law fits over the limited range of pair separations shown,
\begin{equation}
  \xi(r) = \left(\frac{r}{r_0}\right)^\gamma.
  \label{e:xir}
\end{equation}
The fitted values of $r_0$ and $\gamma$ are given in
Table~\ref{t:correlfitdata}.

\begin{table*}
  \begin{center}
    \begin{tabular}{c|c|c|c|c}\hline
      Mass bin  & \multicolumn{2}{c|}{$\lambda<\lambmed$} &
      \multicolumn{2}{c|}{$\lambda\ge\lambmed$} \\
      ($\munit$)  & $\gamma$ & $r_0/(\lunit)$ & $\gamma$ & $r_0/(\lunit)$ \\
      \hline\hline
      $10^{11.4}$--$10^{12}$ & $-1.553\pm0.0013$ & $3.670\pm0.0070$ 
                             & $-1.489\pm0.0012$ & $3.671\pm0.0073$\\
      $10^{12}$--$10^{13}$   & $-1.591\pm0.0031$ & $4.22\pm0.020$ 
                             & $-1.537\pm0.0026$ & $4.82\pm0.021$\\
      $10^{13}$--$10^{14}$   & $-1.64\pm0.021$   & $6.6\pm0.24$ 
                             & $-1.71\pm0.016$   & $8.3\pm0.23$\\
      $10^{14}$--$10^{16}$   & $-1.6\pm0.19$     & $13\pm5.8$ 
                             & $-1.7\pm0.12$     & $19\pm4.8$\\
      \hline
    \end{tabular}
    \caption{Parameters for the power-law $\xi(r)=(r/r_o)^\gamma$ from
    fitting to the eight two-point correlation functions for
    TREEclean haloes.}
    \label{t:correlfitdata}
  \end{center}
\end{table*}

For the higher mass bins, the results from the different types of
haloes are similar: higher spin haloes are more strongly clustered than
lower spin haloes. This could be because objects evolving in denser,
more clustered environments are more likely to experience stronger
tidal forces, leading to more coherent rotation. These results are
consistent with the tentative earlier work of
\cite{1987ApJ...319..575B}, as well as the mark correlation function
analysis of \cite{2002A&A...395....1F}

The difference in clustering strength between high and low spin haloes
decreases with halo mass.  For the least massive TREE haloes,
$M=10^{11.4}$--$10^{12} \munit$, there is virtually no difference in
the correlation functions of the fast and slow rotators.

Note that haloes from the `cleaner' groupfinder (TREE) exhibit
stronger clustering at all scales, for all but the highest masses.  This
is due to the fact that the splitting algorithm applied during the
construction of the merger trees results in a greater number of close
halo neighbours (albeit with reduced masses) compared to the
corresponding FOF haloes.

We have performed a similar analysis to that presented in Figs.
\ref{f:xir_QE_np_TREE} and \ref{f:xir_QE_np_FOF} for the halo
sphericity parameter, $s$.  We found analogous results, in that the
more spherical haloes are more clustered than the less spherical
haloes.

To examine the effect of halo spin and shape on clustering in more
detail, we consider the bias parameter, $b$, which describes how much
more or less clustered a set of haloes is relative to the underlying
dark matter distribution.  We examine how the bias varies for haloes
with different values of $\lambda$ or $s$, at a fixed range of mass.
Similar analyses have recently been performed by
\cite{2005MNRAS.363L..66G}, \cite{2005astro.ph.12416W} and
\cite{2006astro.ph..6699W}, who examined the effect of halo formation
time, concentration, substructure content and time since last major
merger on the bias.

The bias parameter is related to the correlation function through:
\begin{equation}
  \xi_\mathrm{hh}(r|\Mh,\lambda) = b^2(r|\Mh,\lambda)\xi_\mathrm{mm}(r)
  \label{e:bias}
\end{equation}
where $\xi_\mathrm{hh}(r|\Mh,\lambda)$ denotes the halo-halo
correlation function for haloes in a given range of mass and spin (in
this example), and $\xi_\mathrm{mm}(r)$ is the dark matter correlation
function.  We compute the bias parameter as a function of mass,
$b(\Mh)$, using a similar method to that of
\cite{2005MNRAS.363L..66G}.  Specifically, we compute
$\xi_\mathrm{hh}(r|\Mh,\lambda)$ in four $r$-bins in the range $6\leq
r\leq 25\lunit$, equally spaced in $\log r$.  The bias parameter at
each mass is then found as the normalisation constant that minimises:
\begin{equation}
  \chi^2 =\sum_{i=1}^n\left(\frac{\xi_\mathrm{hh}(r_i)-b^2\xi_\mathrm{mm}(r_i)}
                                 {\sigma_\mathrm{hh}(r_i)} \right)^2
  \label{e:biasfit}
\end{equation}
where the $\sigma_\mathrm{hh}(r_i)$ are the Poisson errors on
$\xi_\mathrm{hh}(r_i)$, and the sum is over all ($> 1$) $r$-bins where
$\xi_\mathrm{hh}(r_i)>0$ and $\xi_\mathrm{hh}(r_i)$ was computed using
at least $100$ objects.  This procedure is performed first for all the
TREEclean and FOFclean haloes.  It is then repeated for the haloes in
the upper and lower 20th percentiles of the $\lambda$ and $s$
distributions, from both catalogues.

\begin{figure*}  
  \includegraphics[width=150mm]{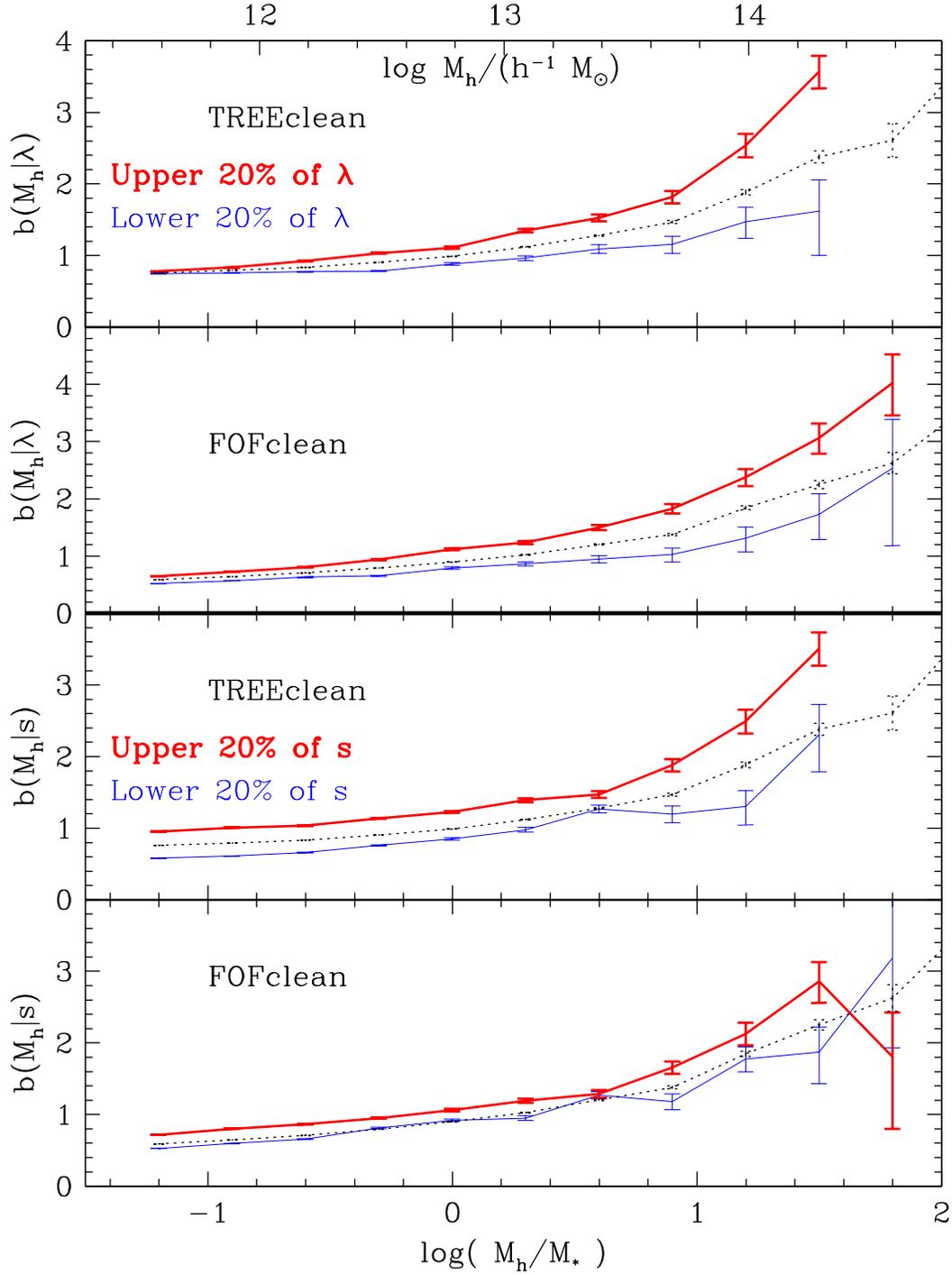}
  \caption{Bias parameter, $b(\Mh)$, for the halo populations
    indicated in the legend.  In each plot, the central dotted line
    marks the bias for the entire population at that mass.  The thick
    red line is the bias for haloes in the upper $20$ per cent of the
    distributions of $\lambda$ (upper two plots) and $s$ (lower two
    plots), and the thin blue line is for haloes in the lower $20$ per
    cent of the distributions.  The error bars give the
    $\Delta\chi^2=1$ confidence interval.  The lines stop either when
    the correlation function $\xi(r)$ for at least $3$ of the $4$
    radial bins is non-positive, or if it was made using fewer than $100$
    objects.}
  \label{f:biasall}
\end{figure*}

The results are shown in Fig. \ref{f:biasall}.  This shows that higher
spin haloes are more clustered than the average, and the lower spin
haloes are less clustered.  This trend is largest at higher masses,
reaching a factor of $\approx 2.2$ between the high-spin and low-spin
bias at the largest mass, $\sim 2\times 10^{14}\munit$.


\section{Conclusions}\label{s:conc}
The huge size and high resolution of the Millennium simulation
\citep{2005Natur.435..629S} makes it possible to determine the properties
of dark matter haloes in the \lcdm cosmology with unprecedented
statistical power. In this paper, we have concentrated on the spins
and shapes of dark matter haloes, ranging in mass from those of dwarf
galaxies to those of rich clusters. We have investigated the
distribution of the spin parameter, $\lambda$, its dependence on mass,
the distribution of shapes, the relationship between shape and spin,
and the environmental dependence of spin and shape. We provide
accurate fitting formulae for several quantities of interest.

While many of the properties we have investigated here have been
studied in earlier simulations going back over twenty years, a novel
aspect of our work is the analysis and comparison of haloes identified
in different ways. Alongside the traditional ``friends-of-friends''
(FOF) algorithm of \cite{1985ApJ...292..371D} and the ``spherical
overdensity'' (SO) algorithm of \cite{1994MNRAS.271..676L}, we have
introduced a new halo definition, the TREE haloes, which are
perhaps the most appropriate when carrying out comparisons of the
simulation results with galaxy and cluster data. The TREE haloes are
defined as branches of the halo merger trees, in which special care has
been taken to identify physical haloes by separating objects that are
artificially and transiently linked together.  Each groupfinder
results in more than $15$ million haloes at $z=0$.

The TREE halo catalogue was further `cleaned' in two ways. Firstly, to
remove any remaining spurious objects, we applied a cut in the
instantaneous virial ratio, ($-0.5 \le \frac{2T}{U}+1 \le
0.5$). Secondly, to remove objects whose angular momentum is biased
due to particle discreteness, we considered only haloes with more than
$300$ particles, as indicated by convergence tests. Our final cleaned
halo catalogues consist of $>10^6$ haloes at $z=0$.

We find that the distribution of the dimensionless spin parameter,
$P(\lambda)$, is poorly fit by a lognormal when this many objects are
considered.  The function given by Eqn.~\ref{e:arjfn} provides a much
better description of the data.  Although the distribution of
$\lambda(\Mh)$ is fairly broad, there is a clear trend of the median
spin with halo mass, with more massive haloes spinning more slowly.
However, the strength and shape of the trend is significantly
different for different halo definitions.  The cubic polynomial of
Eqn.~\ref{e:poly} provides a very good fit to the median spin of the
TREEclean halo catalogue, over a factor of $\sim 10^3$ in halo
mass.

We analysed the shapes of the haloes, and found, as in previous
studies, that there is a broad distribution of shapes with a slight
preference for prolateness over oblateness.  More massive haloes are
less spherical and more prolate in the median, although the data span
a large fraction of the available shape parameter space. We fit broken
lines to the trends with $\log\Mh$ of the median sphericity axis ratio
$s=c/a$ and the median triaxiality parameter $\mathcal{T}$
(Eqn.~\ref{e:broken}).  Both these quantities exhibit a change of
behaviour at a galactic mass scale, $\Mh\approx 2\times 10^{12}\munit$,
where the gradient of the fit changes sign, with haloes becoming
increasingly aspherical and more prolate with increasing mass.

The rounder haloes have less coherent rotation, with a median spin
that is independent of mass ($\lambmed(s\ga0.9)\approx 0.033$). The
most massive haloes have a median spin that is independent of
sphericity ($\lambmed(s)\approx 0.032$).  However, there is
significantly less correlation between the nature of halo triaxiality
(prolateness {\it vs} oblateness) and the spin parameter. Although the
haloes are far from being rotationally supported, there is a strong
preference for the spin vector to be aligned parallel to the halo
minor axis and to be perpendicular to the major axis. The tendency for
the spin to be perpendicular to the intermediate axis is significantly
weakened by the prevalence of prolate shapes for which there is
a near degeneracy between the intermediate and minor axes.

We find a clear signal that the spins and shapes of haloes are
sensitive to the cosmological environment: more rapidly rotating
haloes of a given mass are more strongly clustered. The strength of
this effect increases with halo mass. It is weak for subgalactic and
galactic haloes, but can be larger than a factor of $\sim 2$ for
galaxy cluster haloes.  A similar effect is seen when examining halo
shapes: more spherical haloes are more strongly clustered, with a
greater signal at higher masses.  Our result adds further evidence to
the recent finding by \cite{2005MNRAS.363L..66G}, also from analysis
of the Millennium Run, that the internal properties of haloes depend
not only upon their mass but also upon the environment in which they
form.

The huge number of haloes in the \lcdm Millennium simulation enables
us to characterise the distribution of halo spins, and their relation
to halo mass, shape and clustering, with unprecedented precision.
However, we have also shown the significance of a careful halo
definition.  The properties of haloes defined and identified in
different ways are noticeably different, and it is important to make
the appropriate choice for a given application.  For comparisons with
real data, we recommend using the new class of ``TREE'' haloes which
we have investigated in this work.


\section*{Acknowledgements}
PB acknowledges receipt of a PhD studentship from the Particle Physics
and Astronomy Research Council.  VRE is a Royal Society University
Research Fellow. CSF is a Royal Society-Wolfson Research Merit Award
holder. The simulation and analysis used in this paper was
carried out as part of the programme of the Virgo Consortium on the
Regatta supercomputer of the Computing Centre of the
Max-Planck-Society in Garching, and the Cosmology Machine
supercomputer at the Institute for Computational Cosmology, Durham.

\bibliographystyle{mn2e}
\bibliography{spindistro}  

\appendix   
\section{Examples of Haloes}
The following plots give examples of haloes from our catalogues,
chosen as clear examples of various (usually undesirable) groupfinder
effects -- they are \emph{not} `typical' haloes from their respective
catalogues.  We show them to give some visual intuition as to the
problems that can be encountered with different group-finding
algorithms, as described in section \ref{s:halocats}.

\begin{figure*}
  \includegraphics[width=145mm]{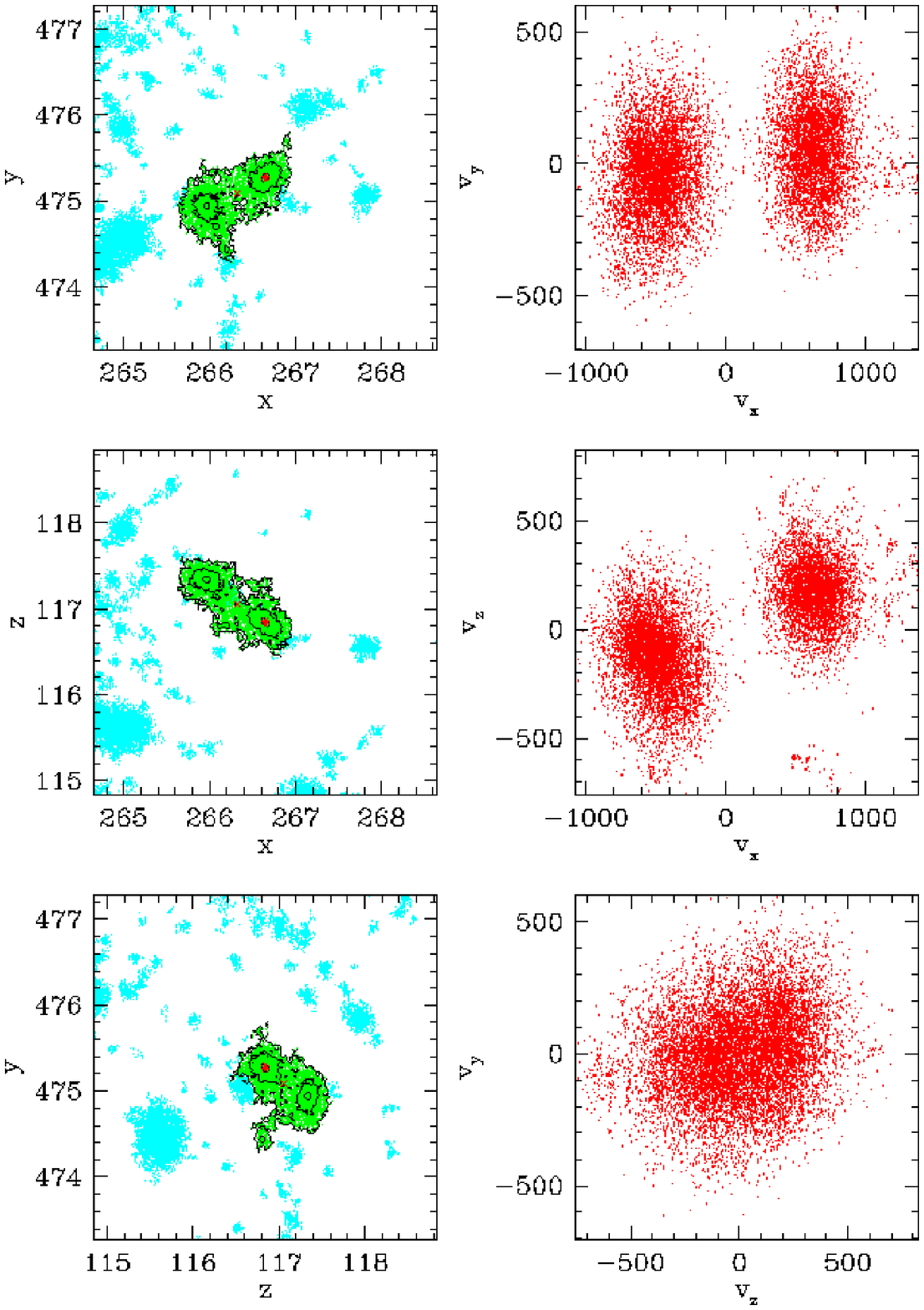}  
  \caption{An example of a FOF halo made of two objects linked by a
   tenuous bridge.  The left-hand panels show projections of the halo
   (overplotted in green) and its neighbours within $2\lunit$ (blue)
   in real space, in units of $\lunit$.  The black contours show
   projected particle density, at $1$, $10$ and $100$ particles per
   contouring bin. The potential-minimum centre of the main halo is
   marked with a red star, and the centre-of-mass is marked with a red
   cross.  The right-hand panels show the particles of the selected
   halo in velocity space, in $\kms$.  This halo has a mass of
   $\Mh=11418\mp\approx 9.82\times 10^{12}\munit$, and a spin
   parameter of $\lambda=1.5712$.  Its virial ratio, $\frac{2T}{U}+1 =
   -4.23$ means it is excluded from the FOFclean catalogue. }
  \label{f:halo_bridging}
\end{figure*}

\begin{figure*}
  \includegraphics[width=145mm]{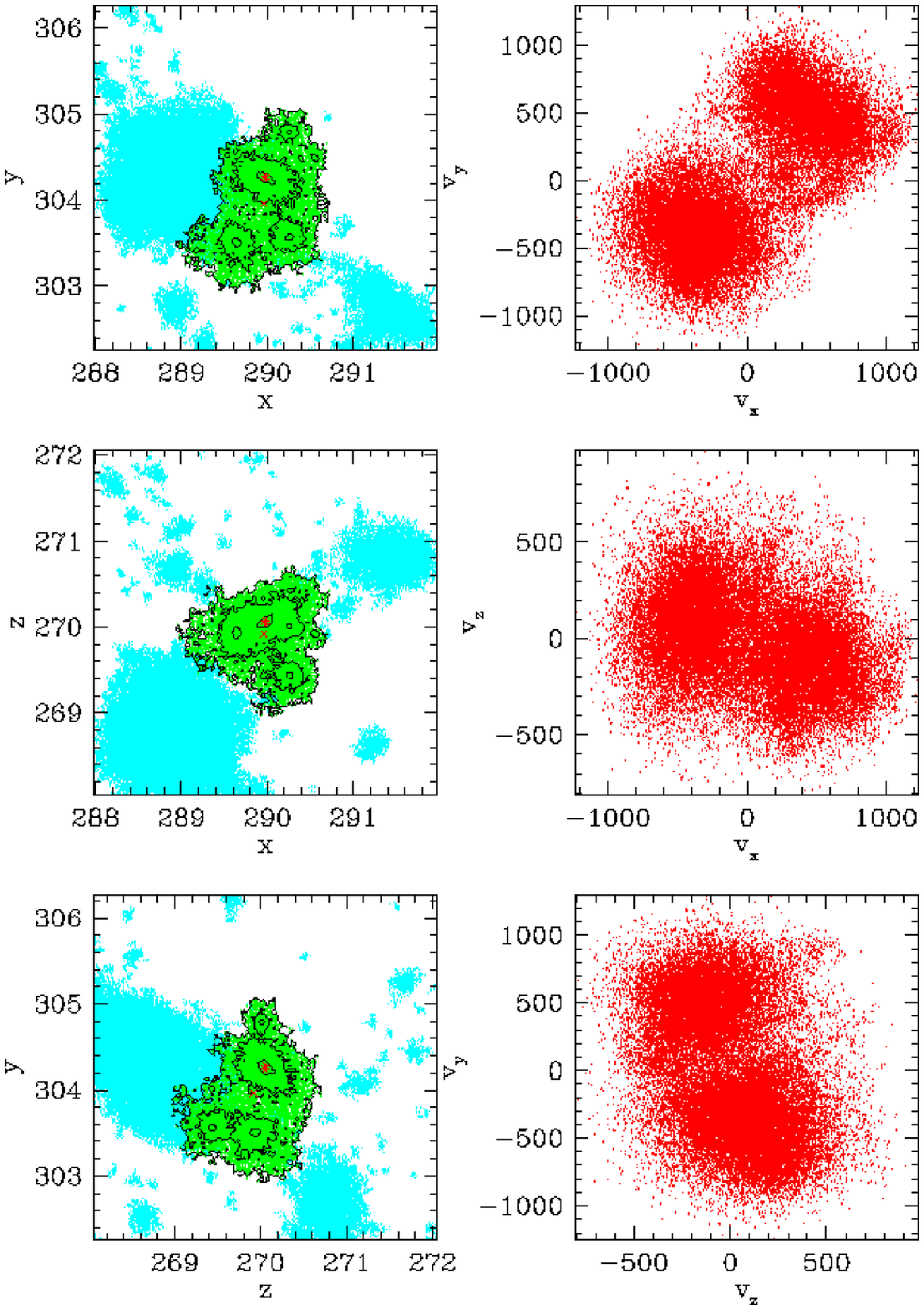} 
  \caption{As Fig. \ref{f:halo_bridging}, but showing an example of
  a larger multi-object FOF halo.  This halo has a mass of
  $\Mh=38\,741\mp\approx 3.33\times 10^{13}\munit$, and a spin parameter
  of $\lambda=0.3295$.  Its virial ratio, $\frac{2T}{U}+1 = -2.05$
  means it is excluded from the FOFclean catalogue. }
  \label{f:halo_largermulti}
\end{figure*}

\begin{figure*}
  \includegraphics[width=145mm]{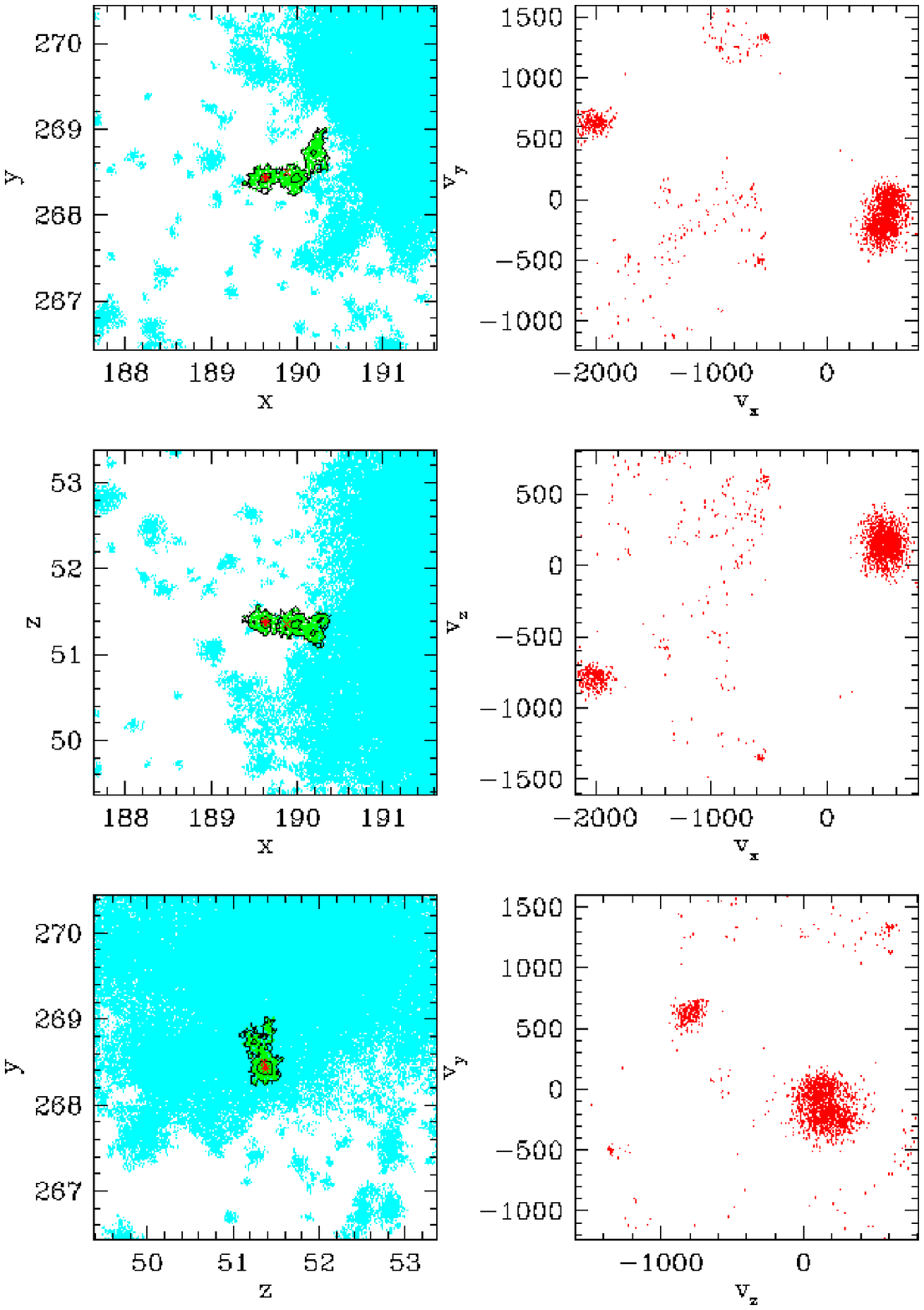} 
  \caption{As Fig. \ref{f:halo_bridging}, but showing an example of
  a small FOF halo with a large neighbour.  The selected halo has a
  mass of $\Mh=1967\mp\approx 1.69\times 10^{12}\munit$, and a spin
  parameter of $\lambda=17.60$. Its virial ratio, $\frac{2T}{U}+1 =
  -53.8$ means it is excluded from the FOFclean catalogue.}
  \label{f:halo_smallneighbour}
\end{figure*}

The first three show FOF haloes with unusual properties in real and
velocity space.  Fig. \ref{f:halo_bridging} shows a halo that is
clearly made up of at least two objects joined via tenuous bridge in.
We show a more massive halo that nevertheless consists of many linked
objects in Fig.  \ref{f:halo_largermulti}.
Fig. \ref{f:halo_smallneighbour} shows a very distorted object located
near a much larger halo.

\begin{figure*}
  \includegraphics[width=145mm]{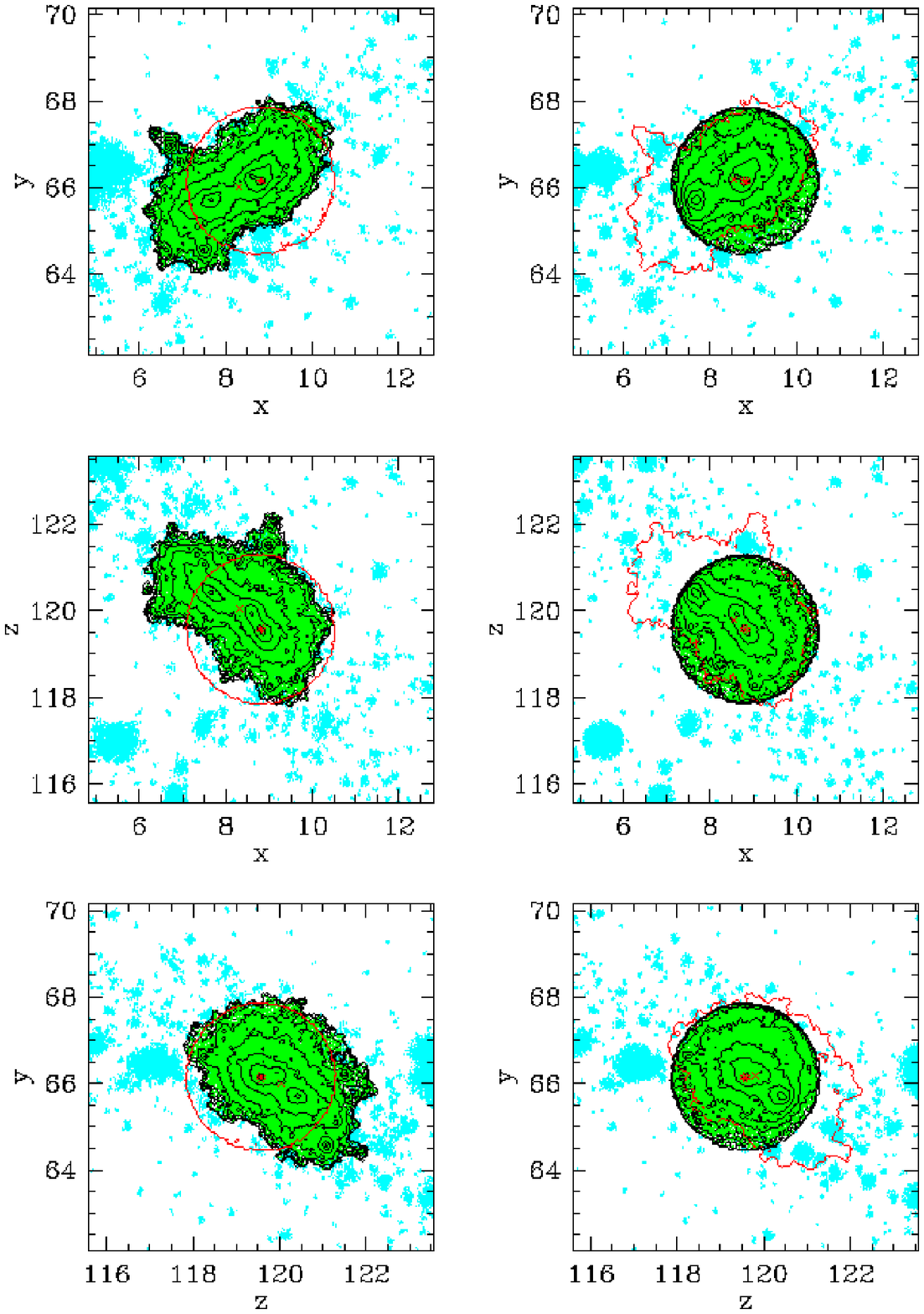}  
  \caption{As Fig. \ref{f:halo_bridging}, but showing a comparison of
  a massive FOF halo (left) and the corresponding SO halo (right), and
  their neighbours in each catalogue within $4\lunit$.  A random
  sample of $1/4$ of the particles are plotted; this doesn't effect
  the overall image of the halo.  Projected logarithmic
  density-contours are plotted in black, spaced every factor of
  $\sqrt{10}$ between $1$ and $10^{3.5}$ particles per contouring bin.
  The outermost contour (the halo boundary) is reproduced on the
  opposite panel in red.  Note how the SO halo includes particles that
  were not part of the FOF group, and are much less dense than the
  halo proper.  The FOF halo has a mass of $\Mh=744\,019\mp\approx
  6.40\times 10^{14}\munit$, and a spin of $\lambda=0.05959$.  It has
  a virial ratio of $\frac{2T}{U}+1 = -0.262$, so it is included in
  the FOFclean catalogue.  The SO halo has a mass of
  $\Mh=610\,023\mp\approx 5.25\times 10^{14}\munit$, and a spin of
  $\lambda=0.04539$. It has a virial ratio of $\frac{2T}{U}+1 =
  -0.332$, slightly less relaxed than its FOF counterpart but still
  included in the SOclean catalogue.}
  \label{f:halo_SObad}
\end{figure*}

\begin{figure*}
  \includegraphics[width=145mm]{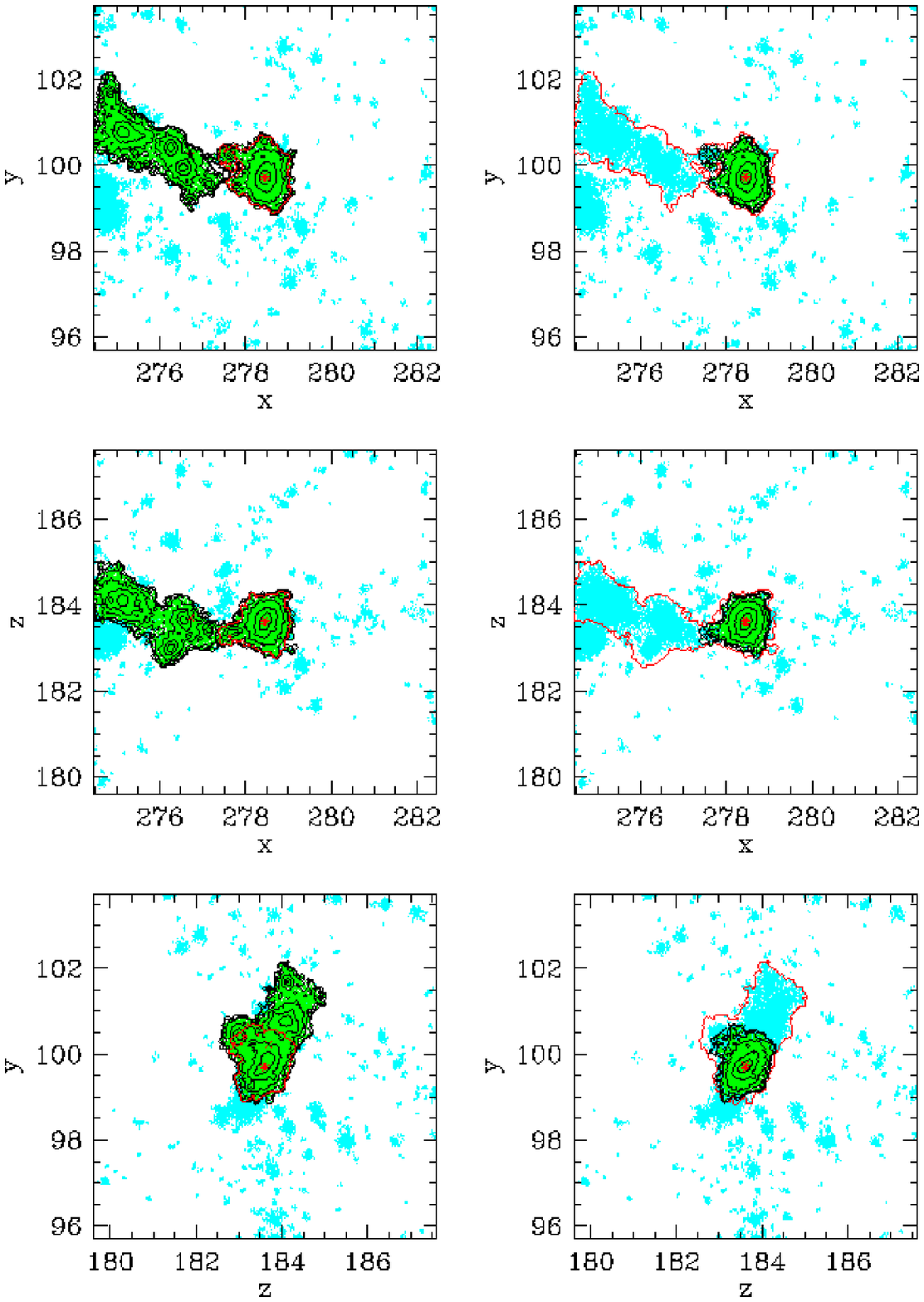}  
  \caption{As Fig. \ref{f:halo_SObad}, but comparing a massive FOF
  halo (left) and the corresponding TREE haloes (right).  Again, a
  random selection of $1/4$ of the points are plotted, and the density
  contours are spaced every factor of $\sqrt{10}$ up to $10^{2.5}$
  particles per contouring bin.  Note how the FOF halo is highly
  extended, with the subhalo housing the potential minimum being
  linked to a large neighbouring halo.  The TREE catalogue splits
  these into two objects.  The FOF halo has a mass of
  $\Mh=126\,033\mp\approx 1.08\times 10^{14}\munit$, and a spin of
  $\lambda=0.1953$.  It has a virial ratio of $\frac{2T}{U}+1 =
  -0.111$, so it is included in the FOFclean catalogue (despite its
  peculiar structure). The TREE halo has a much lower mass of
  $\Mh=40\,719\mp\approx 3.50\times 10^{13}\munit$, and a spin of
  $\lambda=0.05711$. It is slightly less relaxed however, with a
  virial ratio of $\frac{2T}{U}+1 = -0.159$; it is included in the
  TREEclean catalogue.}
  \label{f:halo_TreeGood}
\end{figure*}

The final two figures compare the results of the three different
group-finding algorithms used.  Figs \ref{f:halo_SObad} and
\ref{f:halo_TreeGood} compare haloes defined using the FOF, SO and
TREE algorithms.  In both cases, projections of the selected FOF halo
(and its neighbours) are shown in the left-hand panels, and the
corresponding SO/TREE halo and neighbours are in the right-hand
panels.  The selected haloes (green) have, again, been chosen to
provide a striking illustration of the effects of different
algorithms.  The more `normal' haloes in the background (blue) are
less strongly affected by the choice of groupfinder.

\label{lastpage}
\end{document}